\documentclass[aps,amsmath,amssymb, notitlepage, twocolumn,
footinbib
]{revtex4-1}

\usepackage{graphicx}
\usepackage{color}
\usepackage{bbold}

\newcommand{\psib}{{\boldsymbol\psi}}
\newcommand{\zb}{{\boldsymbol z}}

\newcommand{\bit}{\begin{itemize}}
\newcommand{\eit}{\end{itemize}}

\newcommand{\f}{\frac}
\renewcommand{\>}{\right\rangle}
\newcommand{\<}{\left\langle}
\newcommand{\ba}{\begin{align}}
\newcommand{\ea}{\end{align}}
\newcommand{\be}{\begin{equation}}
\newcommand{\ee}{\end{equation}}
\newcommand{\bi}{\begin{itemize}}
\newcommand{\ei}{\end{itemize}}
\newcommand{\lf}{\left(}
\newcommand{\ri}{\right)}
\newcommand{\dd}{\mathrm{d}}
\newcommand{\DD}{\mathcal{D}}

\newcommand{\sss}{\subsubsection}

\newcommand{\Tr}{\operatorname{Tr}}
\newcommand{\tr}{\operatorname{tr}}
\newcommand{\str}{\operatorname{str}}

\newcommand{\ret}{

}

\begin{document}

\newcommand{\bra}[1]{\< #1 \right|}
\newcommand{\ket}[1]{\left| #1 \>}

\title{Universal statistics of vortex lines}
\author{Adam Nahum and J. T. Chalker}
\affiliation{Theoretical Physics, Oxford University, 1 Keble Road, Oxford OX1 3NP, United Kingdom}
\date{\today}

\begin{abstract}
\noindent
We study the vortex lines that are a feature of many random or disordered three-dimensional systems. These show universal statistical properties on long length scales, and geometrical phase transitions analogous to percolation transitions but in distinct universality classes. The field theories for these problems have not previously been identified, so that while many numerical studies have been performed, a framework for interpreting the results has been lacking. We provide such a framework with mappings to simple supersymmetric models. Our main focus is on vortices in short-range correlated complex fields, which show a geometrical phase transition that we argue is described by the $CP^{k|k}$ model (essentially the $CP^{n-1}$ model in the replica limit $n\rightarrow 1$). This can be seen by mapping a lattice version of the problem to a lattice gauge theory. A related field theory with a noncompact gauge field, the `$NCCP^{k|k}$ model', is a supersymmetric extension of the standard dual theory for the XY transition, and we show that XY duality gives another way to understand the appearance of field theories of this type. The supersymmetric descriptions yield results relevant, for example, to vortices in the XY model and in superfluids, to optical vortices, and to certain models of cosmic strings. A distinct but related field theory, the $RP^{2l|2l}$ model (or the $RP^{n-1}$ model in the limit $n\rightarrow 1$) describes the unoriented vortices which occur for instance in nematic liquid crystals.  Finally, we show that in two dimensions, a lattice gauge theory analogous to that discussed in three dimensions gives a simple way to see the known relation between two-dimensional percolation and the $CP^{k|k}$ sigma model with a $\theta$ term.
\end{abstract}
\maketitle

\section{Introduction}

\noindent
Ensembles of random geometric objects lead to some of the most subtle and interesting questions in critical phenomena. A geometric viewpoint is often useful in analyzing phase transitions: for example, one can view an Ising transition as a proliferation of domain walls, or an XY transition as a proliferation of vortices, leading to important dual descriptions \cite{Kogut review, Jose, banks myerson kogut, peskin, thomas}. But in addition to this, some of the most intruiguing issues concern the critical behaviour of the geometrical objects themselves. In this paper we consider the universal statistical properties of \emph{vortex lines} in random media in three dimensions, a subject which has received considerable numerical attention \cite{tricolour percolation, Vachaspati Vilenkin, Scherrer Frieman, dennis, Strobl Hindmarsh}, but which has so far lacked a field theoretic formulation.

Geometrical critical behaviour can occur even in the absence of a conventional thermodynamic phase transition. The best-known example of this is percolation, in which nonlocal geometrical correlation functions -- such as the probability that two sites lie in the same cluster -- yield universal exponents, despite the fact that correlation functions of local observables remain trivial at the critical point \cite{Essam}. In fact it is precisely because of the short-range correlations of the local degrees of freedom that the percolation universality class is ubiquitous: generic random media are short-range correlated.

Two-dimensional percolation can be thought of as a fluctuating soup of loops \cite{Cardy SLE review}. On the lattice, these are the cluster boundaries; in the continuum, they are the zero lines of a random real height field $v(x)$. When $v(x)$ is biased towards either positive or negative values, the loops have a finite typical size, but these cases are separated by a critical point at which this lengthscale diverges and the loop ensemble becomes conformally invariant.

Vortex lines in three dimensions constitute an analogous problem. The most generic version of it concerns the statistics of the zero lines of a short-range correlated \emph{complex} field $w(x)$. Again, these vortices form a fluctuating soup of oriented loops, with the orientation of a loop given by the sense in which the phase of $w(x)$ rotates around the vortex. Although the correlations of $w(x)$ are trivial by construction, there is an interesting geometrical phase transition as a function of bias -- the ratio of the average value of $w(x)$ to the width of its distribution \cite{tricolour percolation, Strobl Hindmarsh}. As will be explained in detail in the following, there is an \emph{extended} phase at small or zero bias \cite{tricolour percolation, dennis, Vachaspati Vilenkin, Scherrer Frieman}, in which infinite vortex lines proliferate (and are found numerically to be Brownian), while at large bias there is a \emph{localized} phase in which long vortices are exponentially suppressed. In between lies a critical point, where loops have a nontrivial fractal structure (and a fractal dimension close to $5/2$). Similar transitions can also occur in the absence of applied bias, for example in the vicinity of the critical point of the XY model, or for line defects in other order parameters: in particular, one can consider the unoriented ($\mathbb{Z}_2$) vortices in nematics \cite{kibble Z2, Strobl Hindmarsh}.

Universal aspects of vortex statistics have been examined numerically in a wide range of contexts, including cosmic strings, beginning with the simulations of Vachaspati and Vilenkin \cite{Vachaspati Vilenkin, Strobl Hindmarsh, Scherrer Frieman, Scherrer Vilenkin, kibble Z2, Vachaspati}, line defects in random light fields, known as `optical vortices' \cite{dennis, dennis topology}, vortices in the XY \cite{Kajantjie, Bittner} and Abelian Higgs models \cite{Wenzel, Hove, Prokofev Svistunov, Vincent}, in turbulent superfluids \cite{superfluid turbulence}, and in a Potts-like model introduced to describe polymers \cite{tricolour percolation}. In addition, similar scaling behaviour has been observed in simulations of lattice models of completely-packed loops \cite{ortuno somoza chalker, Jaubert}.

Despite this previous work on vortices, numerous questions remain unanswered, because of the lack of a continuum formulation in which the relevant correlation functions can be expressed. In particular, it has not been clear how to classify these problems, or what the criteria are for two instances to be in the same universality class. It has been uncertain, for example, whether critical exponents for the geometrical vortex transition coincide with those for conventional three-dimensional (3D) percolation, or whether universal behaviour for oriented and unoriented vortices coincide (even in the extended phase), or how the geometrical transition for vortices in the XY model relates to the thermodynamic one. Even basic facts such as the appearance of Brownian statistics in the extended phase have not been derived, though previously this has been explained heuristically \cite{Scherrer Frieman} by analogy with polymers in the melt, long known to be Brownian \cite{polymers in melt}.

Our effort to address these questions starts from a set of new duality transformations. We begin by showing that the partition function for tricolour percolation, a Potts-like lattice model used for the numerical investigation of vortex statistics \cite{tricolour percolation, Scherrer Frieman, Strobl Hindmarsh}, can be mapped exactly to that of a supersymmetric lattice gauge theory via a high temperature (graphical) expansion of the latter. By considering the continuum limit of this lattice gauge theory, and, separately, by discussing mappings from vortex problems that are formulated directly in the continuum, we arrive at field theories describing the geometrical properties of vortices.

For oriented vortices, these field theories involve a complex supervector $\psib=(z^1,\ldots z^{k+1}, \chi^1,\ldots \chi^k)$, with $k+1$ bosonic and $k$ fermionic components, coupled to a $U(1)$ gauge field $A$. (Supersymmetry \cite{McKane, Parisi Sourlas SUSY for loops} is a standard tool in problems involving loops or polymers, allowing  representation of geometrical correlation functions.) In order to describe the transition for vortices in short-range correlated complex fields $w(x)$, the gauge field $A$ must be taken to be compact, leading to a model in the universality class of the $CP^{k|k}$ sigma model \cite{read saleur}. Heuristically, this can be thought of as the replica limit $n\rightarrow 1$ of the $CP^{n-1}$ sigma model \cite{di vecchia, Witten CPn-1}, which at $n=2$ is equivalent to the classical Heisenberg (or $O(3)$) model. The ordering transition in the sigma model describes the geometrical transition in the vortex ensemble. $CP^{k|k}$ field theories have been related to lattice models of completely packed loops in both 2D -- by Read and Saleur \cite{read saleur} and Candu, Jacobsen, Read and Saleur \cite{candu et al} -- and 3D \cite{short loop paper, forthcoming long version}. (Strictly speaking we are abusing terminology, since there is a \emph{separate} $CP^{k|k}$ model for each integer $k\geq 0$, the model with $k=0$ being trivial. The correspondence with vortices holds for arbitrary $k$, but the range of geometrical correlation functions which can be expressed in the theory grows with $k$.)

The $CP^{k|k}$ description provides the appropriate framework for questions about vortex statistics. Immediate consequences include the explanation for Brownian statistics in the extended phase, which corresponds to the ordered phase in the sigma model language, and qualitative features of the critical point (for example, the fact that the fractal dimension is close to $d_f=5/2$ is easily understood as a symptom of confinement in the gauge theory). Most importantly, it clarifies the relations between the different universality classes, allowing us to classify these problems and explaining why various numerically studied models are in the same universality class and others are not. For instance, it explains the coincidence of exponents measured for the bias induced transition for vortices \cite{tricolour percolation} and for the completely packed 3D loop models \cite{ortuno somoza chalker, cardy class C review}, which also permit a mapping to the $CP^{k|k}$ model \cite{short loop paper}.

In the absence of bias in the distribution for $w(x)$, the gauge field in the supersymmetric theory is non-compact,  yielding the non-compact $CP^{k|k}$ or $NCCP^{k|k}$ model, to borrow the terminology used for non-compact $CP^{n-1}$ models. The appearance of this theory can be simply understood by reference to a well-known duality which maps the XY model to an Abelian Higgs theory with a complex scalar field $z$ coupled to a gauge field \cite{banks myerson kogut, thomas, peskin, dasgupta halperin, fisher XY duality, kleinert XY duality, zee}. Under this mapping, XY vortices become worldlines of the particles created by $z^*$ (to use a $2+1$ dimensional language). However, this standard dual description does not allow us to probe the long distance behaviour of vortices, because it does not allow us to write down the required geometric correlation functions. The $NCCP^{k|k}$ model is a supersymmetric extension of the Abelian Higgs model with a richer operator content (and possibly a richer phase diagram).

Unoriented vortices (e.g. in a short-range correlated random nematic order parameter) are described by a different field theory, the $RP^{2l|2l}$ model. This is essentially the replica limit of the $RP^{n-1}$ sigma model. A soft spin version of the $CP^{n-1}$ model in the replica limit allows us to clarify the relation between these two geometric problems, as well as the relation to 3D percolation, and also shows that the upper critical dimension for $CP^{k|k}$ is six, meaning that exponents can be calculated in a $6-\epsilon$ expansion \cite{Amit, McKane Wallace Zia, Bonfim, Senthil UCD}. The $CP^{1|1}$ model also describes Anderson localization in symmetry class C, which is closely related to classical loop models \cite{mirlin evers review, glr, beamond cardy chalker, ortuno somoza chalker, cardy class C review}, so we expect six to be an upper critical dimension for this problem too. A similar conclusion has been reached previously by Senthil \cite{Senthil UCD}, who considered the soft spin formulation of $CP^{1|1}$.

As mentioned above, another important conceptual connection is with 2D percolation. One of the continuum descriptions of this problem is as a $CP^{k|k}$ model with $\theta$ term, as was shown via a mapping to a supersymmetric spin chain \cite{read saleur, candu et al}. To connect this up with our 3D results, we give a treatment of 2D percolation as a lattice gauge theory (with an unconventional Boltzmann weight for the gauge field) analogous to our treatment of tricolour percolation in 3D. This leads in a simple way to the continuum description. A similar lattice regularization of $CP^{n-1}$ has been discussed by Affleck \cite{Affleck CPn-1 and SOS}.

The structure of the paper is as follows. We begin in Sec. \ref{overview} with a heuristic overview of the key ideas. In Sec. \ref{short range correlations} we give duality mappings for oriented vortices in short-range-correlated random systems on the lattice (\ref{tricolour percolation section}) and in the continuum (\ref{vortex derivation}). For the lattice derivation, we use a specific model, tricolour percolation \cite{tricolour percolation, Scherrer Frieman}, an elegant realization of the vortex problem which may be mapped to a lattice gauge theory. Sec. \ref{statistics of vortex lines} gives the basic consequences of the field theory for vortex statistics.

Sec.~\ref{Other universality classes} considers variations of the vortex problem. First we briefly describe the soft spin version of the $CP^{n-1}$ model in the limit $n\rightarrow 1$ and the $6-\epsilon$ expansion (Sec.~\ref{upper critical dimension}). We then consider a perturbation leading to crossover to conventional percolation (Sec.~\ref{crossover to percolation}). Here we also discuss the issue of vortex intersections which is important in many realistic vortex problems. Sec.~\ref{unoriented vortices} considers unoriented vortices, occurring when the fundamental group of the order parameter manifold is $\mathbb{Z}_2$. Sec.~\ref{long range correlations} considers the stability of the universal behaviour of vortices in a short-range correlated $w(x)$ to the introduction of long-range correlations. We then consider vortices near the (thermodynamic) critical point of the XY model, their description in terms of the $NCCP^{k|k}$ model (Sec.~\ref{XY critical point}), and the phase structure of this field theory. This concludes our discussion of vortices. In Sec.~\ref{percolation} we consider loop models in 2D and their relation to lattice gauge theories with $\theta$ terms. We close with a short summary in Sec.~\ref{Summary}.

\section{Heuristic considerations}
\label{overview}

\noindent
We begin with a handwaving argument as to the form of the field theory for vortices in a short-range-correlated random complex function $w(x)$. This is intended to provide an overview and to supplement the formal derivations of Sec. \ref{short range correlations}.

To reiterate, we expect the following phase structure. When the probability distribution for $w(x)$ is $U(1)$ symmetric, or only weakly biased, vortices have a fractal dimension of two and a finite fraction of the vortex density is in infinite vortex lines. This is the extended phase. A nonzero (translationally invariant) mean value suppresses vortices; when the mean is sufficiently large compared with the typical size of fluctuations, we enter the localized phase, and infinite vortices disappear \cite{tricolour percolation, Vachaspati}. At the continuous transition between these phases, vortices are random fractals.

In general, a simple way to relate ensembles of loops to field theory is to view the loops as worldlines of quantum particles -- in this case in 2+1 dimensions -- and to ask what sort of particles these must be and what kind of interactions they must have. The partition function for the field theory is a sum over histories of the quantum system: expressed in the right basis, this is a weighted sum over worldline configurations. The Lagrangian must be chosen so that this sum reproduces the weighted sum over loop configurations we started with.

Of course, to be useful the correspondence must go beyond the partition function: we should be able to represent geometrical correlation functions -- such as the probability that two points $x$ and $y$ lie on the same loop, denoted $G(x-y)$ -- as correlation functions of local operators in the field theory. The field theory we seek must allow us to do this despite the fact that $G(x-y)$ is nonlocal in the original description, in the sense that we cannot necessarily tell, by looking only in the vicinity of $x$ and $y$, whether the two points lie on the same loop.

\begin{figure}[b] 
\centering
\includegraphics[width=1.5in]{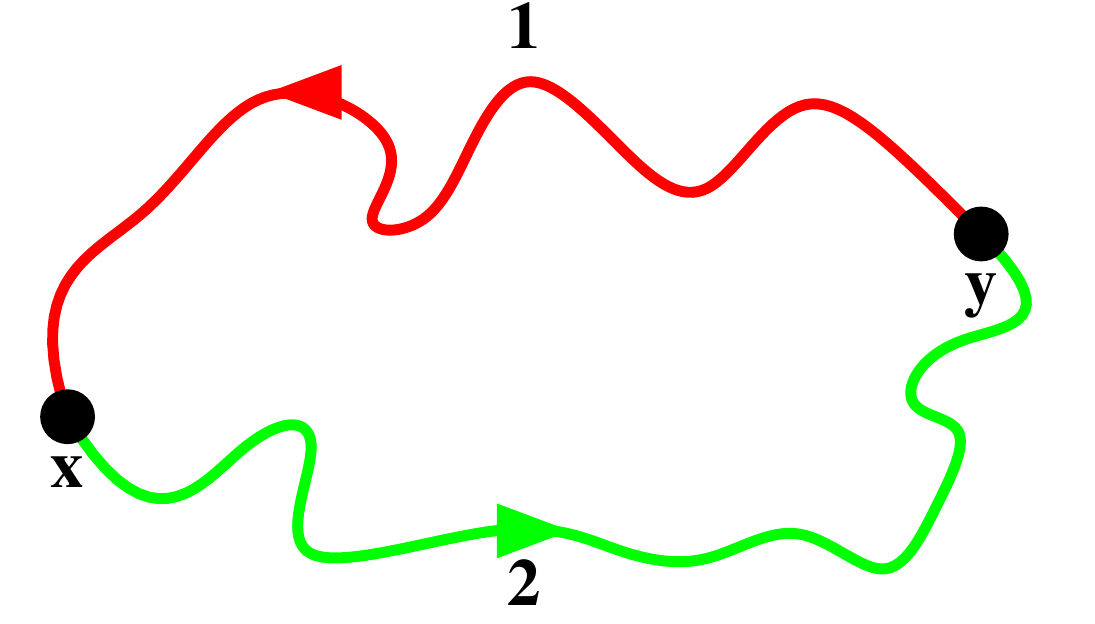}\hspace{5mm}
\includegraphics[width=1.5in]{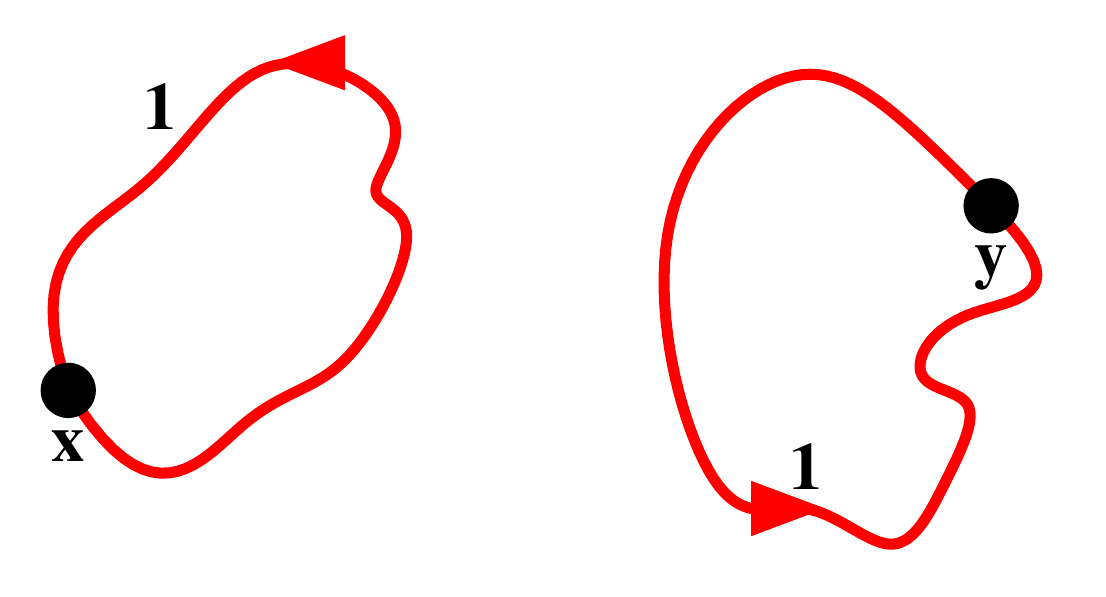}
\caption{(Color online) Left: The two-point function  $\<z^1 z^{2*} (x) z^2 z^{1*}(y) \>$ forces $x$ and $y$ to be joined by a loop. With only a single colour index, we could not separate such configurations from those with the topology shown on the right.}
\label{correlatorpic}
\end{figure}

Since vortices have an orientation determined by the sense of rotation of $\arg w(x)$, it would be natural to assume that we will need a single species of charged boson -- with vortices that go forward and backward in the imaginary time direction corresponding to worldlines of particles and antiparticles respectively. As usual, these bosons would be represented by a single complex scalar field $z(x)$. However, this construction would not allow us to write correlators like $G(x-y)$. To be able to do that, we must give the worldlines (and particles) an additional colour label $\alpha=1,...,n$, so that $z$ expands to a vector $z^\alpha$. This yields operators such as $z^1(x) z^{2*} (x)$, which, when inserted into a correlation function, forces a meeting at $x$ between an outgoing worldline strand of colour 2 and an incoming strand of colour 1. Therefore, neglecting for now the possibility of strands escaping to infinity, the two point function  $\<z^1 z^{2*} (x) z^2 z^{1*}(y) \>$ forces $x$ and $y$ to be joined by a loop, one arm of which is of colour 1, and the other of colour 2 (see Fig.~\ref{correlatorpic}). This is just what we need to construct $G(x-y)$. By contrast, if we had only a single species of particle at our disposal, we could not separate out configurations with a loop joining $x$ and $y$ from those with $x$ and $y$ lying on separate loops.

But while the colours solve this problem, they introduce another one. The sum over colour indices for each loop modifies the weight for a given configuration of loops by the unwanted factor $n^{\text{no. loops}}$: we are no longer describing the problem of vortices, but a completely different problem where loops have a fugacity equal to $n$. 

One way to get around this is to take a replica-like limit $n\rightarrow 1$ at the end of any calculation \cite{replica for loops}. A more concrete alternative \cite{McKane, Parisi Sourlas SUSY for loops, read saleur} is to exchange $z$ for a supervector
\be
\psib = (z^1,...,z^{k+1},\chi^1,...\chi^k)
\ee 
with $k+1$ bosonic and $k$ fermionic components. Fermionic loops come with a minus sign. Thus, summing over the possible colours for a loop yields a fugacity of $k+1-k =1$ as desired, independently of $k$. 

What should the Lagrangian for $\psib$ be? It must be invariant under supersymmetry transformations preserving $\psib^\dag \psib$. At a technical level, this ensures that the partition function is independent of the number of fermions $k$. In terms of loops, the invariance is a consequence of the fact that the colour index we have introduced is just a label: the weight of a configuration does not change if we alter the colour of one of the loops (except by minus signs associated with fermionic colours). It is tempting also to assume that since $w(x)$ is short-range correlated, interactions between vortices will be short-range, and therefore that we should write down a field theory with local interactions. The natural candidate is 
\be
\label{wrong lagrangian}
\mathcal{L}_\text{incorrect} = |\nabla \psib|^2 + \mu |\psib|^2 + \lambda |\psib|^4.
\ee
However, this is not what we want. The phase transition (as a function of $\mu$) at which $\psib$ condenses does indeed represent a proliferation of $\psib$ worldlines. But this phase transition is thermodynamically nontrivial -- by setting $k=0$ we see it is in the XY universality class -- and so cannot represent the vortex phase transition we are discussing, where all local degrees of freedom are short-range correlated.  

As a hint towards a better candidate, notice that correlation functions of the form
\be
\< [\psi(x)]^{N_\text{in}} [\psi^*(x)]^{N_\text{out}}  \lf\ldots\ri\>,
\ee
where $\lf\ldots\ri$ contains insertions away from $x$ (and we omit colour indices) have no meaning in the vortex problem unless $N_\text{in} =N_\text{out}$: for obvious topological reasons, the number of vortex strands entering the vicinity of $x$ must equal the number leaving. We are thus free to set these correlation functions to zero, which can be done by introducing a gauge symmetry:
\be
\label{Lagrangian with A and ...}
\mathcal{L} = |(\nabla - i A) \psib|^2 + \mu |\psib|^2 + \lambda |\psib|^4 +  \ldots
\ee
The significance of this change depends on what Lagrangian for the gauge field is hidden in the ellipsis in (\ref{Lagrangian with A and ...}). In the trivial limit of zero gauge coupling, when non-gauge fluctuations of $A$ are completely suppressed, the thermodynamics of (\ref{wrong lagrangian}) will be unchanged and we will not have made any progress. On the other hand, a nontrivial action for $A$ will mediate interactions between particles which depend on their charge. This is what we want -- it reflects the fact that the weight of a given configuration of vortices depends on their relative orientations, as is clear from thinking about the winding of $\arg w(x)$.

Our previous assumption of local interactions for $\psib$ was in fact incorrect. The weight for a vortex configuration $\mathcal{C}$ comes from integrating over all compatible configurations of $w$, and this yields a long-range interaction between vortices. The role of the integral over $A$ is to simulate this part of the statistical weight. This is the fundamental idea of XY duality \cite{banks myerson kogut, thomas, peskin, dasgupta halperin, fisher XY duality, kleinert XY duality, zee, samuel quark confinement} -- in the context of the present discussion, see especially \cite{thomas, kleinert XY duality}. The coupling of loops to a gauge field has also been exploited to derive exact results for self-avoiding loops in two dimensions \cite{Cardy mean area, Cardy scaling functions}.

Specifically, in the presence of the gauge field each configuration $\mathcal{C}$ of oriented loops is weighted by the expectation value 
\be
W (\mathcal{C}) = \< \exp i \int_\mathcal{C} A. \dd x \>_A,
\ee
of a Wilson loop calculated using the functional integral over $A$ only, and it is the desired form of this weight which determines the required action for $A$. It is particularly easy to see what this should be in the limit of large bias. Suppose that $\<w\>$ is large and positive, and consider $W(\mathcal{C})$ for a single large loop. Then $\arg w(x)$ will be close to zero for most $x$, but the vortex must be the boundary of a sheet where $\arg w(x) = \pi$. This sheet will cost an `energy' scaling as the minimal area $\mathcal{A}$ enclosed by $\mathcal{C}$, so that $\log W(\mathcal{C}) \sim - \mathcal{A}$.

Such an area law for the Wilson loop implies that $A$ should be a compact gauge field \cite{polyakov compact gauge theory, Polyakov later paper, kleinert XY duality}: we must include singular Dirac monopole configurations in the functional integral over $A$. While we used large bias for this discussion, monopoles are in fact present at any nonzero bias. (Since they are irrelevant in the extended phase, the zero-bias point is not distinguished from the rest of the extended phase as far as the universal statistics of vortices are concerned.)

The field theory we have arrived at, with a supervector coupled to a compact gauge field $A$ as in Eq.~(\ref{Lagrangian with A and ...}), will appear naturally in our derivations. However it is often more convenient to use alternative formulations which describe the same universal physics. We now discuss two such formulations. We temporarily simplify the discussion by exchanging $\psib$ for $n$-component bosonic vector $\zb$, with the replica limit $n \rightarrow 1$ in mind. 

For the first, we switch to a sigma model, with the constraint  $\zb^\dag \zb =1$ rather than a potential for $\zb$, and the Lagrangian:
\ba
\label{CPn-1 lagrangian without gauge field}
\mathcal{L}_{CP^{n-1}} &= \f{1}{g^2} \lf | \nabla \zb|^2 - |\zb^\dag \nabla \zb|^2 \ri, &
\zb^\dag \zb & = 1.
\end{align}
The peculiar-looking kinetic term is designed to be insensitive to the phase of $\zb$, so that we retain the gauge symmetry $\zb \rightarrow e^{i \phi} \zb$. Because of this gauge symmetry, the local degrees of freedom live on $CP^{n-1}$, the manifold of $n$-component complex unit vectors $z$ subject to the identification $\zb \sim e^{i\phi} \zb$ \cite{di vecchia, Witten CPn-1}. The sigma model describes the ordering transition of `spins' living on this space. At long distances it is equivalent to the compact gauge theory, in which the effect of the gauge field is to confine the charged particles $\zb$ into neutral composite particles, or alternately to render the overall phase of $\zb$ a redundant degree of freedom.

\begin{figure}[h] 
\centering
\includegraphics[height=0.55in]{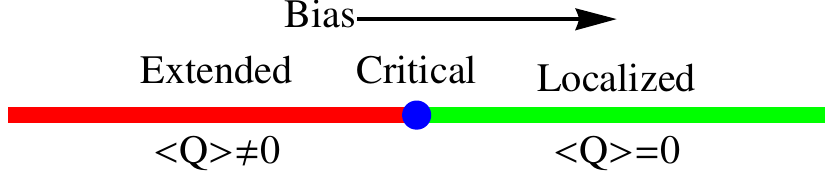}
\caption{(Color online) Phase diagram for vortices in short-range correlated $w(x)$.}
\label{1Dphasediagram}
\end{figure}

The order parameter for the transition in this sigma model is a gauge-invariant traceless $n\times n$ matrix, 
\ba
Q & = \zb \zb^\dag - (\zb^\dag \zb) / n, &
\tr Q & = 0.
\end{align}
In the limit $n\rightarrow 1$, or in the supersymmetric formulation described below, the transition is that between localized and extended vortex phases, and the extended phase is that in which $\<Q\>\neq 0$. The phase diagram is shown in Fig.~\ref{1Dphasediagram}.

When $n=2$, the order-disorder transition is simply that of the $O(3)$ model, to which the $CP^1$ model is equivalent. This equivalence is seen by using the Pauli matrices to form a three-component Heisenberg spin, $S^i  = \tr \sigma^i Q = \zb^\dag \sigma^i \zb$: when written in terms of $S$, the sigma model Lagrangian (\ref{CPn-1 lagrangian without gauge field}) becomes that of the $O(3)$ sigma model. The fact that $n=1$ (the `$CP^0$' model) must be regarded as a limiting case is seen in the fact that $Q=0$ in this limit.

A familiar idea in the context of the $O(3)$ model is that a soft spin formulation in which $S^2$ is unconstrained is often more useful than a sigma model in which $S^2 =1$. The second formulation we will need is an analogous soft spin theory for $Q$,
\be
\label{first appearance of soft spin theory}
\mathcal{L}_\text{soft spin} = \tr \, (\nabla Q)^2 + \mu \tr Q^2 + \nu \tr Q^3 + \ldots
\ee
(the cubic term vanishes in the special case $n=2$). This formulation, in the $n\rightarrow 1$ limit, will be useful to us in Sec.~\ref{Other universality classes} in considering the effect of perturbations to the vortex problem. An immediate inference from it is that the upper critical dimension for this universality class (and for the related problem of the Anderson transition in class C \cite{Senthil UCD}) is six. Exponents for vortices can thus be computed in a $6-\epsilon$ expansion similar to that for percolation \cite{Amit, McKane Wallace Zia, Bonfim}.

The supersymmetric replacement for the replica limit of the sigma model (\ref{CPn-1 lagrangian without gauge field}) is got simply by replacing $\zb$ with $\psib$,
\ba
\label{sigma model lagrangian}
\mathcal{L}_{CP^{k|k}} &=  \f{1}{g^2} \lf | \nabla \psib |^2 - |\psib^\dag \nabla \psib |^2 \ri, &
\psib^\dag \psib & = 1.
\end{align}
In two dimensions, these supersymmetric sigma models and their connection with loop models have been studied in \cite{read saleur, candu et al}. The gauge-invariant operator $Q$ becomes a supermatrix \cite{Efetov},
\ba
\label{first appearance of Q}
Q &= \psib \psib^\dag -(\psib^\dag \psib), & \str Q &= 0.
\end{align}
The transition into the ordered (extended) phase as a function of the coupling $g$ is equivalent to the spontaneous breaking of supersymmetry and the appearance of an expectation value $\<Q\>$.

This simple phase diagram can be extended in various ways. We can for example modify the problem so as to induce a crossover to the universality class of conventional percolation, or to that for nematic vortices (Sec.~\ref{Other universality classes}). The  gauge theory Eq.~(\ref{Lagrangian with A and ...}) also prompts us to ask about the geometrical critical behaviour described by the version with a \emph{noncompact} gauge field. Adopting the standard terminology for the case without fermions \cite{NCCPn-1}, we will call this the noncompact $CP^{k|k}$ model, or $NCCP^{k|k}$. Since compactness (the presence of Dirac monopoles) is a consequence of bias, the $NCCP^{k|k}$ model describes the case without an applied bias. It is a supersymmetric extension of the well-known dual theory for the XY transition \cite{banks myerson kogut, thomas, peskin, dasgupta halperin, fisher XY duality, kleinert XY duality, zee}, and is relevant to the statistics of vortices in the XY model at zero magnetic field \cite{Kajantjie, Bittner}. Here we regard $w(x)$ as the order parameter for an XY model, and we are no longer necessarily in the domain of short-range correlations. We will consider this noncompact theory in Sec.~\ref{XY critical point}. An added complication, in comparison to the compact case, is that it may show two separate phase transitions, one thermodynamic (associated with the onset of the Higgs mechanism) and one geometrical (associated with the breaking of supersymmetry). This is suggested by numerical results for XY vortices \cite{Kajantjie, Bittner}. Surprisingly, in this scenario the geometrical transition may again be described by a \emph{compact} $CP^{k|k}$ model.

\section{Vortices in a short-range correlated random environment}

\label{short range correlations}

\noindent
We now turn to derivations of the $CP^{k|k}$ model for vortices, beginning (Sec.~\ref{tricolour percolation section}) with a lattice model in which geometrical observables can be explicitly related to correlation functions of $\psib$. We then present a derivation directly in the continuum in order to emphasize the universality of the results and to see the relation with more standard treatments of duality (Sec. \ref{vortex derivation}). Sec.~\ref{statistics of vortex lines} then gives simple consequences of the field theory description.

\subsection{Tricolour percolation}
\label{tricolour percolation section}

\noindent
Tricolour percolation is an elegant discretization of the vortex problem \cite{tricolour percolation, Scherrer Frieman} which has become a standard setup for numerical simulations, partly because some more obvious alternatives to this model suffer from aesthetic deficiencies. For example, while it would be natural to consider random phases $\theta \equiv \arg w$ on the sites of a cubic lattice, with vortex lines piercing the plaquettes, this requires an additional rule for deciding when a plaquette is pierced by a vortex line. Also, in such a model nothing prevents vortex lines from intersecting each other, requiring a further decision about how to resolve the intersections. For a generic function $w(x)$ in the continuum on the other hand, zero lines intersect with probability zero. (Intersections will be discussed in detail in \ref{crossover to percolation}.)

Tricolour percolation avoids these issues by a smart choice of lattice, and by discretizing not only real space but also the target space for the degrees of freedom. This idea is familiar from 2D percolation, where a single universality class encompasses both continuum percolation, which deals with zero lines of a continuous random function $v(x)$, and lattice site percolation, where the function takes only the values $v = \pm 1$, and where  intersection of cluster boundaries is avoided by using sites on the triangular rather than the square lattice. For a random complex function, the minimal discretization is to let $w$ run over the cube roots of unity, $w = 1$, $e^{2\pi i /3}$, $e^{4 \pi i /3}$ \cite{Vachaspati Vilenkin}. The configurations are therefore configurations in percolation with three colours \cite{tricolour percolation}, and  vortices (or tricords) are lines where all three colours meet. 

The construction described below was introduced by Scherrer and Frieman \cite{Scherrer Frieman} as an improvement on the model of Vachaspati and Vilenkin \cite{Vachaspati Vilenkin} and continues to be used for numerics in the context of cosmic strings \cite{Strobl Hindmarsh}. It was  introduced independently  and simulated extensively by Bradley, Debierre and Strenski \cite{tricolour percolation}, who realized that the model displays a novel geometric phase transition at nonzero bias. Their interest came from a different direction: they were looking for an efficient means of generating configurations of self-avoiding polymers, and were inspired by earlier work on smart walks \cite{smart walks}.

\subsubsection{Model}

\begin{figure}[b] 
\centering
\includegraphics[width=1.55in]{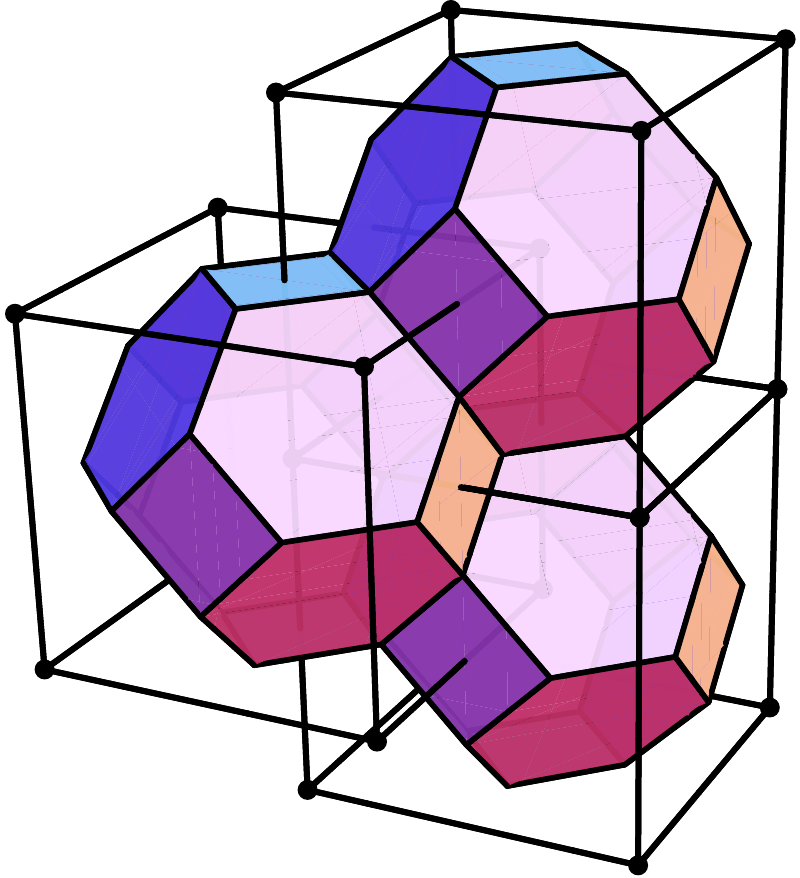}
\includegraphics[width=1.55in]{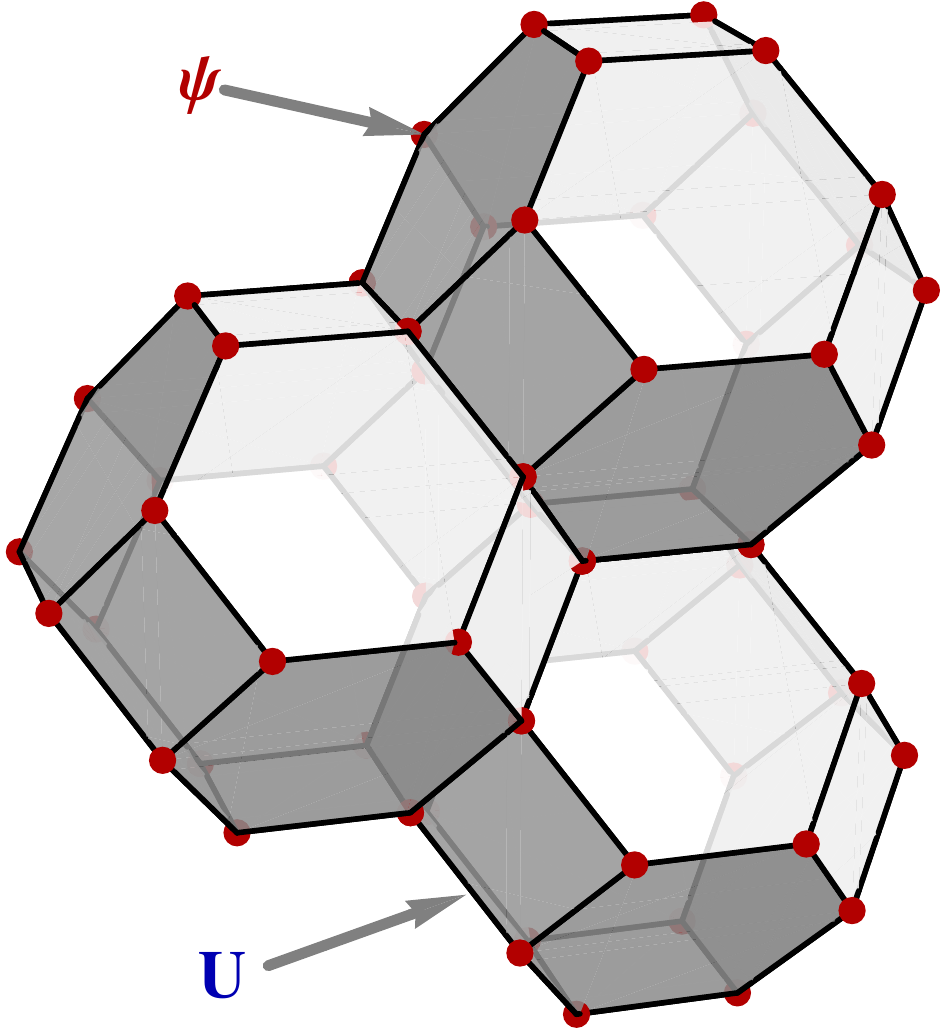}
\caption{(Color online) The lattice for tricolour percolation. Left: showing the relation to the bcc lattice. Right: with the locations of the dual degrees of freedom ($\psib$ on sites and $U$ on links) indicated.}
\label{tp lattice}
\end{figure}

\begin{figure}[b] 
\centering
\includegraphics[height=1.55in]{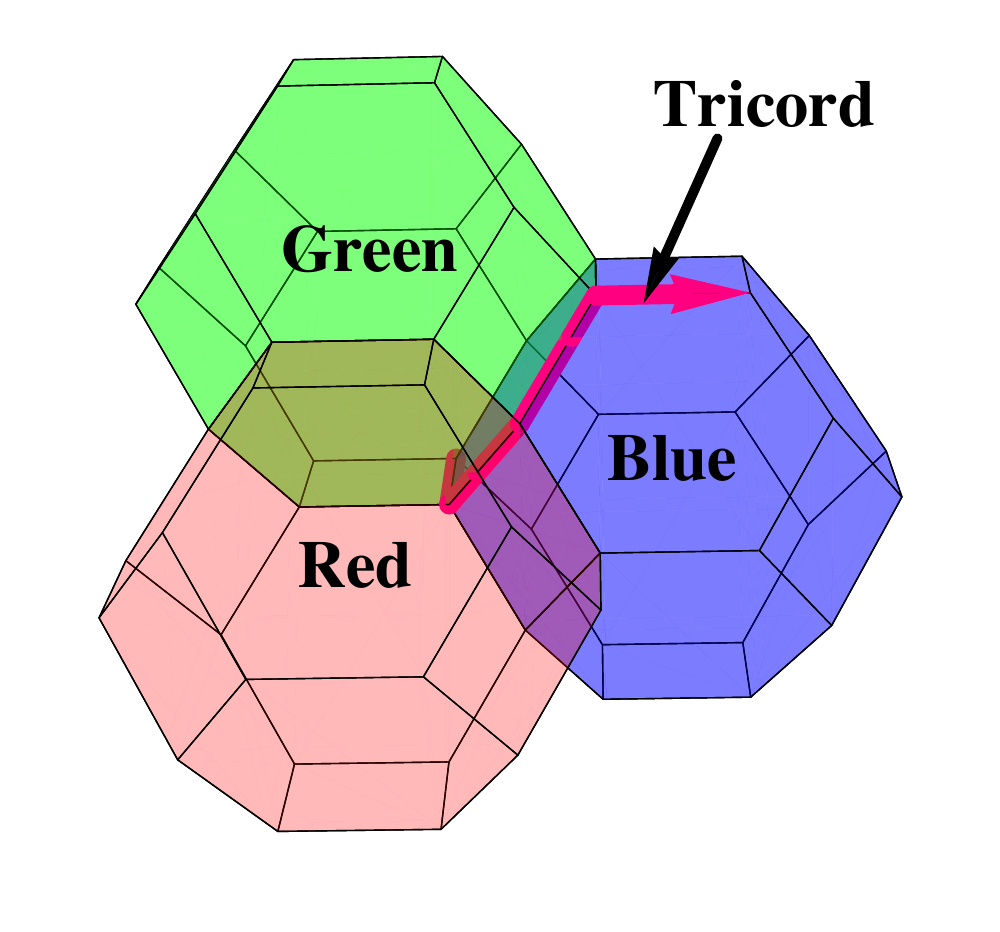}
\caption{(Color online) A section of tricord in tricolour percolation.}
\label{section of tricord}
\end{figure}

\noindent
Fig.~\ref{tp lattice} depicts a tiling of space with cells of a certain shape (truncated octahedra). The centres of these cells lie on a bcc lattice, and they are the Wigner-Seitz cells for this lattice. The loops will live on the edges of the cells, which form a second lattice known as the tetrakaidekahedral lattice. Its key features are that each link is surrounded by three cells, and that its sites are four-coordinated. 

Tricolour percolation configurations are generated by randomly and independently colouring the cells either red, green or blue, with probabilities $p_\text{R}$, $p_\text{G}$ and $p_\text{B}$. Assigning each colour to a cube root of unity
\ba
\{{\bf R, G,B}\}&\longrightarrow\{1,\omega,\omega^2\}, &\omega&=e^{2\pi i/3},
\end{align} 
gives the relation with a discrete random field $w(x)$; the threefold anisotropy in its distribution is not expected to change the universality class of the transition (Sec.~\ref{vortex derivation}). Vortices are the lines along which {all three colours meet}. The design of the lattice ensures that these curves are non-branching and self and mutually avoiding, and they have an orientation defined by the order in which the three colours encircle the link.

Close to the symmetric point, $p_\text{R}=p_\text{G}=p_\text{B}=1/3$, tricords are in the extended phase \cite{tricolour percolation, Strobl Hindmarsh}. Moving sufficiently far from this point, we enter the localized phase.  Numerical studies of the transition have been performed only on the line $p_\text{R} \neq p_\text{G} = p_\text{B}$, which intersects the critical line at the two points $p_\text{R}  =  0.417(1)$ and $p_\text{R} =0.255(5)$ \cite{tricolour percolation}, but the derivation below leads us to believe that the critical behaviour will be the same at all points on the critical line, as expected by the authors of \cite{tricolour percolation}. Note that the critical line lies \emph{inside} the region in which all three colours percolate, so the transitions do not coincide with percolation transitions of the colours (the critical probability for site percolation on the bcc lattice is $0.2460(3)$ \cite{tricolour percolation}).

\subsubsection{Symmetric point}
\label{Symmetric point}

\noindent
We begin with the symmetric point, where the mapping to lattice gauge theory is simplest. The gauge theory we will need is unconventional in that its Boltzmann weight is not naturally written as the exponential of something simple. This situation is standard in the study of loop ensembles: lattice magnets with Boltzmann weights of non-exponential form are useful because they give straightforward graphical expansions, which can be interpreted as loop models \cite{domany}. For a pedagogical introduction to these ideas, see \cite{cardy book}. The motivation here is similar: a graphical expansion of the kind standard in lattice gauge theory \cite{Kogut review} will yield tricolour percolation configurations. 

Let $i$, $j\dots$ label sites of the tetrakaidekahedral lattice, and let $\<ij\>$ denote a link directed from $i$ to $j$. We introduce gauge fields $U_{ij}$ on the links, which are unimodular numbers with $U_{ij}=U_{ji}^*$, and unit supervectors \cite{McKane, Parisi Sourlas SUSY for loops}
\ba
\psib_i&=(z^1_i,...,z^{k+1}_i,\chi_i^1,...,\chi^k_i), & 
\psib^\dag_i \psib_i & =1
\end{align}
at sites. `$\Tr$' will denote an integral over all degrees of freedom, with the normalization $\Tr 1 = 1$. The normalization of $\psib$ then implies that $\Tr \psi_i^\alpha\psi_i^{*\beta}=\delta^{\alpha\beta}$.

Letting $F$ denote a face (where two cells meet) and $l$ a link, the partition function for the lattice gauge theory is
\ba
\notag
Z
=
\Tr \prod_{F}  & \lf   1 +  \prod_{l\in F} U_l  + \prod_{l\in F} U_l^*  \ri \\
& \quad\quad \quad \times
\prod_{\<ij\>} \lf 1 +  U_{ij}^3 \psib^\dag_i \psib^{\phantom{\dag}}_j
+  \text{c.c.} \ri.
\label{LGT for 3-perc}
\end{align}
The links in the product $\prod_{l\in F}$ are oriented consistently around the face $F$, and changing the orientation corresponds to exchanging $\prod U$ and $\prod U^*$.

\begin{figure}[b] 
\centering
\includegraphics[width=1.05in]{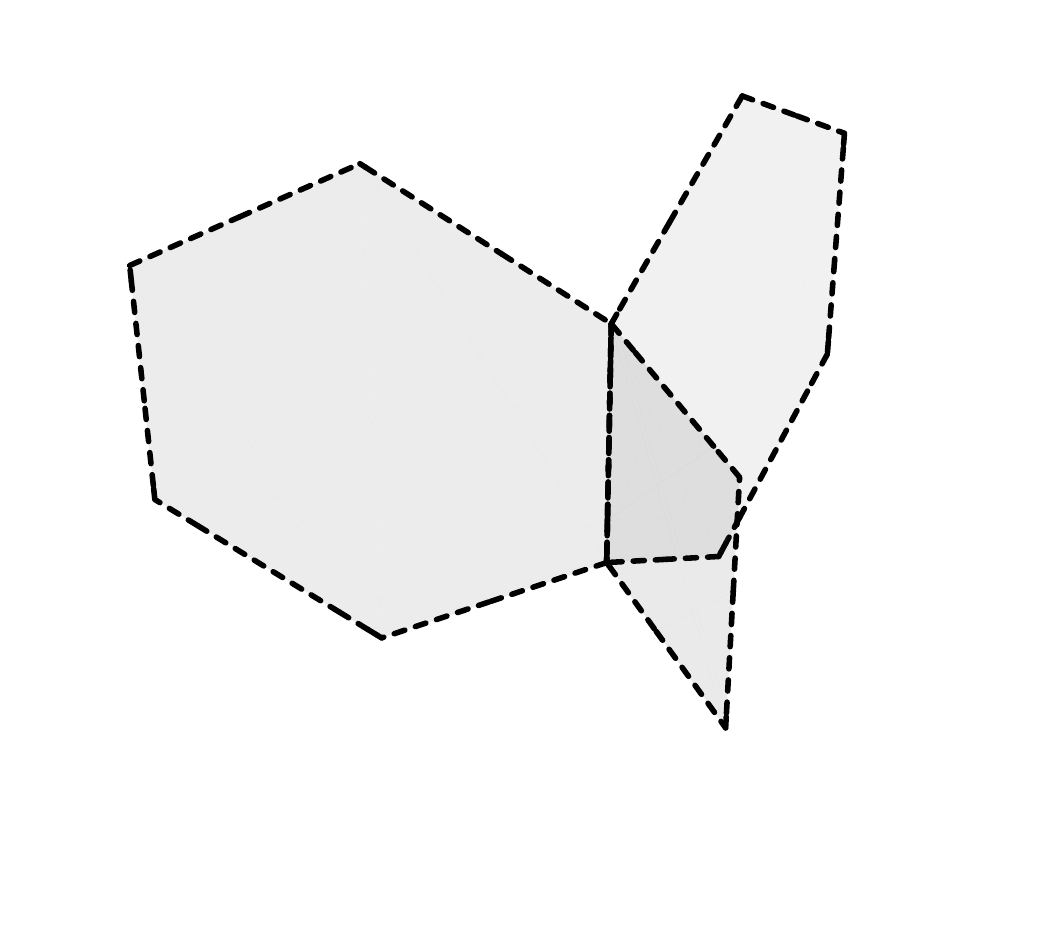}\hskip0.1in
\includegraphics[width=1.05in]{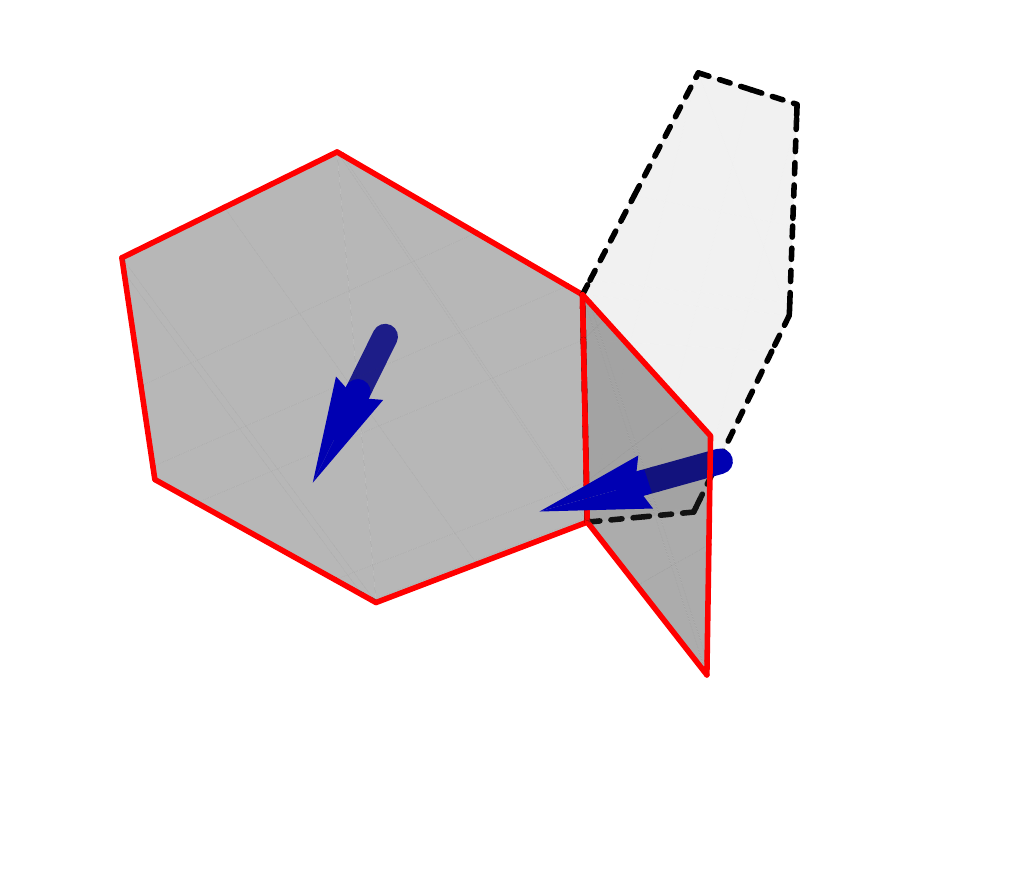}\hskip0.1in
\includegraphics[width=1.05in]{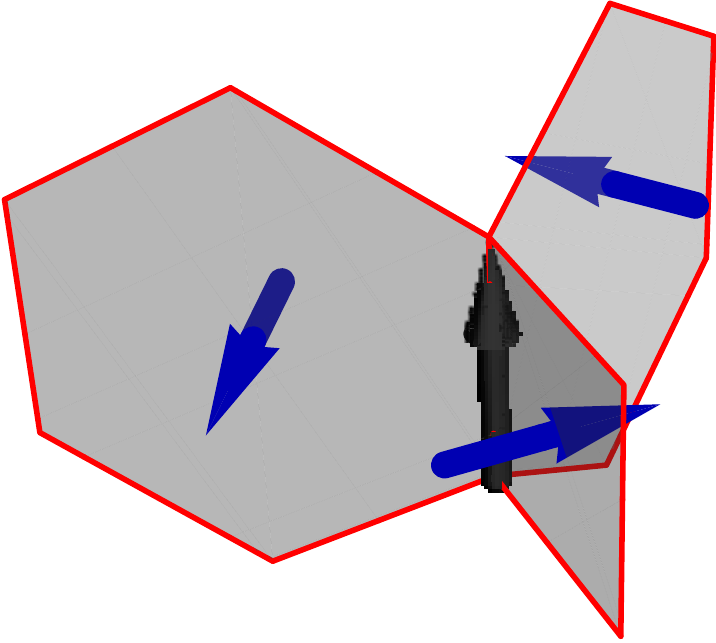}
\caption{(Color online) Possibilities for the graphical expansion at a link, as described in the text. In the mapping to tricolour percolation, the shaded surfaces become domain walls between different colours.}
\label{graphexp}
\end{figure}

To see the relation to tricolour percolation, we expand out the two products in Eq.~(\ref{LGT for 3-perc}) -- one over faces $F$ and one over links $\<ij\>$ -- and represent the terms graphically. The diagram for a given term is built up as follows. For each face $F$, we must choose either $1$, $\prod U$ or $\prod U^*$. If we choose `1' we add nothing to the diagram, while if we choose $\prod U$ or $\prod U^*$ we draw in the face $F$ together with an orientation (equal or opposite to that of the links in the product respectively). We represent this orientation by a normal vector as in Fig.~\ref{graphexp}. For each link we must choose either $1$, in which case we draw nothing, or $U_{ij}^3\psib_i^\dag \psib_j$, in which case we draw in the link with an arrow from $j$ to $i$, or $U_{ji}^3\psib_j^\dag \psib_i$, in which case we orient the arrow the other way.

Only a subset of the diagrams generated this way survive after integrating over $U$, namely those having an equal number of $U_l$ and $U_l^*$ on each link $l$. For each link there are three possibilities (shown in Fig.~\ref{graphexp}). Either (i) none of the three faces bordered by the link is included in the diagram, and neither is the link; (ii) two of the faces are included, they are consistently oriented so as to form part of a sheet of oriented surface, and the link is not included; or (iii) the link is included and so are all three of the faces, with  orientations determined by that of the link and the right hand rule.
We are then left with sheets of oriented surface which close on themselves or meet along directed lines $\mathcal{L}$. On a finite lattice, these lines are closed loops. They are self and mutually avoiding, and neither branch nor terminate. 

Such configurations are easily mapped to tricolour percolation by regarding the sheets as boundaries between domains of different colour. Fixing the colour of one cell, we can colour all the others by the rule that the colour changes cyclically,
\be
{\bf R}\rightarrow{\bf G}\rightarrow{\bf B}\rightarrow{\bf R},
\ee
upon crossing a sheet in the same direction as its normal. The lines $\mathcal{L}$ where three sheets meet are then the tricords. A convenient convention is to consider a finite bcc lattice and define the colour \emph{outside} the lattice to be (say) red. Thus a cell on the boundary is red if its exterior faces (those on the boundary) are not covered by surface in the graphical expansion.

Let $\mathcal{T}$ denote a tricolour percolation configuration with loops $\mathcal{L}$. At this point, having integrated over $U$ but not $\psib$, we have
\be
\label{tricolor Z with psi}
Z= 
\sum_\mathcal{T} \left\{
\Tr \prod_{\mathcal{L} \in\mathcal{T}} (\psib_1^\dag \psib^{\phantom{\dag}}_2) .... (\psib_L^\dag \psib^{\phantom{\dag}}_1) \right\},
\ee
where $1,\ldots, L$ label the sites on a given loop $\mathcal{L}$. To evaluate the integral over the supervectors, note that each loop gives
\be
\Tr \lf \psi_1^{\alpha*} (\psi_2^\alpha \psi_2^{*\beta}) (\psi_3^\beta \psi_3^{*\gamma}) \ldots
(\psi_L^{\psi}\psi_L^{*\omega})
\psi_1^\omega \ri.
\ee
The formula $\Tr \psi^\alpha_i \psi^{\beta*}_i = \delta^{\alpha\beta}$ requires all colour indices to be equal, say to $\alpha$, to give a nonzero result. For the fields $\psi_1$ and $\psi_1^*$, however, a reordering is necessary in order to use this formula: this gives a minus sign when $\alpha$ corresponds to a fermionic component of $\psib$. Thus the sum on $\alpha$ yields a fugacity per loop of $k+1 - k=1$, and $Z$ becomes as a sum over equally weighted tricolour percolation configurations, $Z = \sum_\mathcal{T}1=3^{\text{no. cells}}$. 

In addition, we can use $\psib$s to construct geometrical correlation functions. Let the operators $O_i$ depend only on the $\psib$s. Then the graphical expansion goes through as before and we have
\be
\label{tricolour perc correlator formula}
\notag
\<O_1\ldots O_N \> = \f{1}{Z} 
\sum_\mathcal{T}
\Tr O_1\ldots O_N
\prod_{\mathcal{L} \in\mathcal{T}} (\psib_1^\dag \psib^{\phantom{\dag}}_2) .... (\psib_L^\dag \psib^{\phantom{\dag}}_1).
\ee
This formula allows us to translate correlators of $\psib$s into the probabilities of various geometrically-defined events. The other input required is a formula for the single site traces, 
\ba\notag
\Tr &  \lf \psi^{\alpha_1}\ldots\psi^{\alpha_q} 
\psi^{*\beta_1}\ldots\psi^{*\beta_q} \ri   \\
&\qquad\quad\quad=\f{1}{q!} \< \psi^{\alpha_1}\ldots\psi^{\alpha_q}
\psi^{*\beta_1}\ldots\psi^{*\beta_q} \>_\mathrm{G},
\label{single link traces}
\end{align}
where $\<\ldots\>_\mathrm{G}$ is a Gaussian average evaluated using Wick's theorem and $\< \psi^{\alpha}
\psi^{*\beta}\>_\mathrm{G}=\delta^{\alpha\beta}$.  A simple example is the two-point function $\< Q^{12}_i Q^{21}_j\>$ of the gauge-invariant operator $Q_i^{12}=z_i^1 z_i^{*2}$. The operator insertions restrict the sum over $\mathcal{T}$ to configurations in which $i$ and $j$ lie on the same tricord, and so gives the probability of this event. Note that in order to construct this correlation function we require $k\geq 1$. We will return to correlation functions in Sec. \ref{statistics of vortex lines}.

The naive continuum limit of the lattice gauge theory (\ref{LGT for 3-perc}) involves a superfield $\psib$ coupled to a gauge field $A$. The compactness of the microscopic gauge degrees of freedom $U_l$ translates in the continuum into the presence of Dirac monopoles in $A$ \cite{polyakov compact gauge theory, Polyakov later paper}. However, because $\psib$ couples to $U^3$ rather than to $U$ in (\ref{LGT for 3-perc}), these monopoles are of three times the minimal charge allowed by the Dirac quantization condition. (We will take the minimal value allowed by the Dirac quantization condition as our unit of monopole charge -- hence the factor of three in Eq.~\ref{tricolour perc magnetic field} below.)

Crudely, monopoles arise because there is a $6\pi$ ambiguity in the magnetic flux $B_F$ passing through each face $F$ -- here $F$ denotes a face together with a choice of orientation -- and thus also in the divergence of this quantity. $B_F$ is a vector field on the links of the dual lattice, which pierce the faces $F$, and is defined modulo $6\pi$ by
\be
\label{tricolour perc magnetic field}
\exp\lf \f{i B_F}{3 }\ri = \prod_{l \in F} U_l.
\ee
The divergence $(\nabla.B)_\mathcal{B}$ is defined at each site of the dual lattice, or equivalently each tricolour percolation cell $\mathcal{B}$, by the sum of $B_F$ over the (outwardly oriented) faces of the cell.

Recall that an ambiguity of this kind can be resolved in two ways, which correspond to thinking about $B$ either as a vector field whose divergence is nonzero at monopole defects, or as a vector field with strictly zero divergence but with Dirac strings. If we  resolve the ambiguity by restricting $B_F$ to the domain $[-3\pi, 3\pi)$, it can have nonzero divergence at the locations of monopoles, whose charge  $\rho$ is a multiple of three. Letting $\mathcal{B}$ denote a site of the dual lattice, or equivalently a tricolour percolation cell, 
\ba
(\nabla.B)_\mathcal{B} & = 2 \pi \rho_\mathcal{B}, & \rho_\mathcal{B} & \in 3 \mathbb{Z}.
\end{align}
Alternatively we can resolve the ambiguity (call the new version $B'$) in such a way that
\be
(\nabla.B')_\mathcal{B}  = 0.
\ee
$B'$ differs from $B$ by the inclusion of Dirac strings, which carry $6 \pi$ flux away from the monopoles along strings of adjacent plaquettes, and on which $B'$ lies outside the region $[-3\pi, 3\pi)$.

In a continuum formulation, the functional integral over $A$ must include  configurations with pointlike monopoles (of charge $\rho \in 3\mathbb{Z}$), together with singular Dirac strings which do not incur any weight in the action \cite{Polyakov later paper, Kleinert compact gauge fields}. A fugacity will also be associated with each monopole. Anticipating Sec.~\ref{vortex derivation}, these charge three monopoles are a result of threefold anisotropy in the argument of our random field $\arg w$ \cite{kleinert XY duality}. We postpone further discussion until Sec. \ref{vortex derivation}, since they will not play an important role. The symmetric point lies in the phase where $\psib$ is condensed and $A$ is massive by the Higgs mechanism; the appropriate description here is in terms of Goldstone modes associated with the breaking of supersymmetry, and the monopoles are irrelevant \footnote{Charge-three monopoles would play an important role in a transition into the localized phase, at zero bias, induced by Potts-like interactions between the colours of the cells (these can be included by modifying the Boltzmann weight for the gauge field of Eq. \ref{LGT for 3-perc}).}. However, monopoles of charge one appear on moving away from the symmetric point, and will play an important role at the transition.

\subsubsection{Alternative view of the graphical expansion}
\label{Alternative view on graphical expansion}

\noindent
For an alternative view on the graphical expansion, and to connect up with the discussion in Sec.~\ref{overview}, we can expand out only the second product (that over links) in the partition function (\ref{LGT for 3-perc}). The diagrams now consist only of oriented links (to each of which is attached a factor $U^3\psib^\dag \psib$). Next, integrating over $\psib$ kills all configurations except those in which the links form closed loops. However it does not prevent these loops from intersecting. If $\mathcal{C}$ denotes such a configuration of loops, then
\ba\notag
Z &=  \sum_\mathcal{C} \Tr_U 
\prod_{F}  \left(   1 +  \prod_{l\in F} U_l  + \prod_{l\in F} U_l^*  \right)
\prod_{\text{loops }\mathcal{L}} \lf  \prod_{l \in \mathcal{L}} U^3_l \ri.
\end{align}
In the worldline picture, we see that each loop configuration is weighted by the expectation value of a Wilson loop $W(\mathcal{C})$,
\ba
\label{Wilson loop tp}
Z&\propto 
\sum_\mathcal{C}  W(\mathcal{C}),
&
W(\mathcal{C}) &= \< \prod_{\mathcal{L}\in\mathcal{C}} \prod_{l\in \mathcal{L} } U_l^3 \>_U,
\end{align}
evaluated using the $U$-dependent part of the Boltzmann weight:
\be
\< \ldots \>_U = \f{1}{Z_U} \Tr \, (\ldots) \prod_{F}  \lf   1 +  \prod_{l\in F} U_l  + \prod_{l\in F} U_l^* \ri.
\ee
The expectation value $W(\mathcal{C})$ assigns the correct entropic weight to each tricord configuration. This is a nontrivial weight given by summing over the possible domain wall configurations compatible with a given set of tricords. In particular the weight is zero for loop configurations which are not allowable tricord configurations, including those with intersections of loops.

\subsubsection{Away from the symmetric point}
\label{Away from symmetric point}

\begin{figure}[b] 
\centering
\includegraphics[height=1.8in]{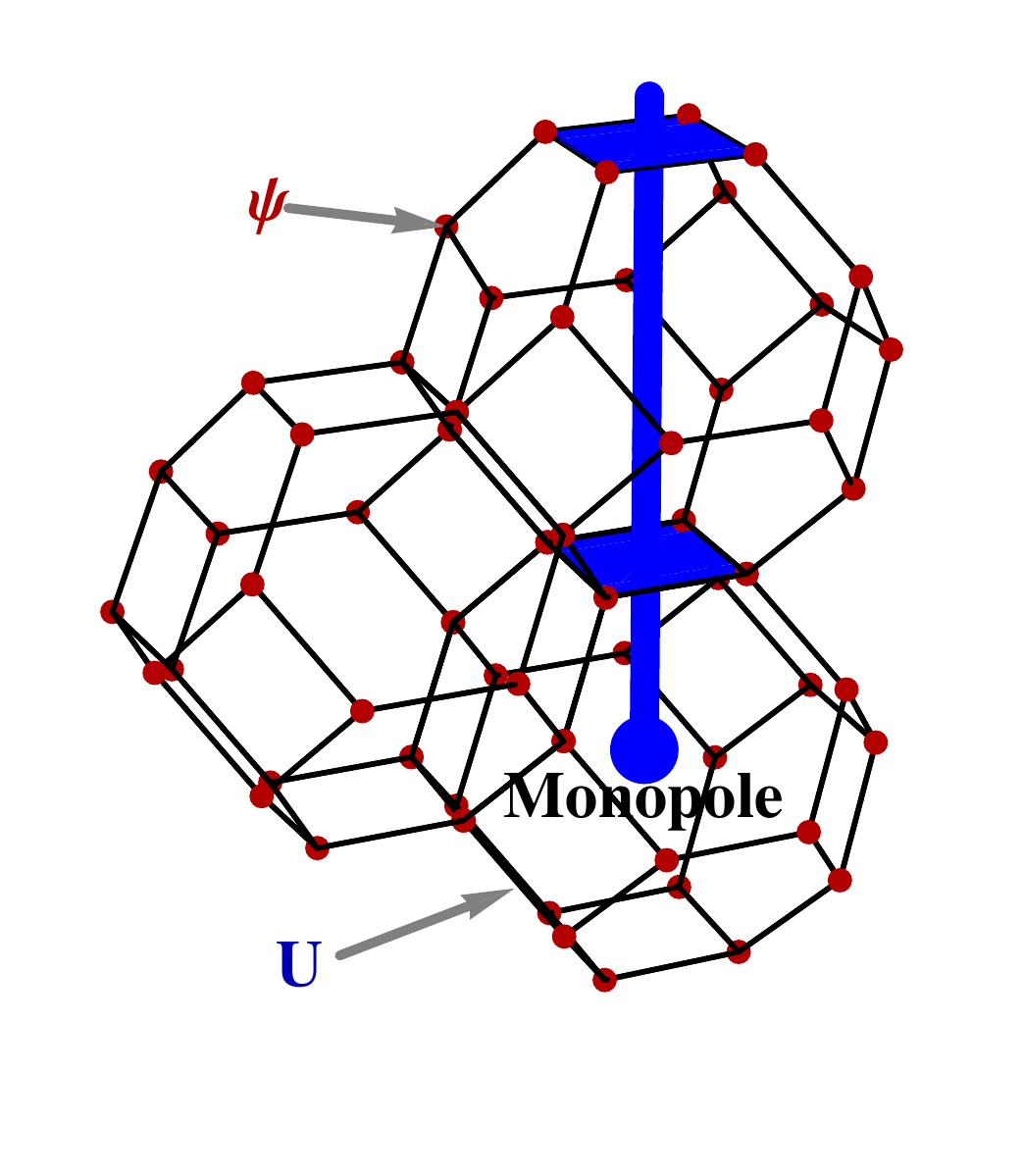}
\caption{(Color online) Away from the symmetric point, monopoles of unit charge appear. In the presence of monopoles, the Boltzmann weight for the gauge field is modified on the plaquettes crossed by Dirac strings.}
\label{TPmonopole}
\end{figure}

\noindent
At first sight, extending the correspondence between tricolour percolation and lattice gauge theory away from the symmetric point presents a problem, since the basic objects in the graphical expansion of the lattice gauge theory are the domain walls between colours rather than the colours themselves. But in fact this is not the case. To begin with we consider changing the probabilities for a single cell: this allows the appearance of a Dirac monopole of unit charge at the centre of that cell. 

Again, we consider a system with boundary, and stipulate that the colour \emph{outside} the boundary is ${\bf R}$. The colour of an interior cell $\mathcal{B}$ is then determined by the signed  number of domain walls we must cross (modulo three)  in order to reach a point outside the boundary, starting from $\mathcal{B}$.

Pick any path $\mathcal{P}$ from the interior of $\mathcal{B}$ to the boundary. Now introduce a variable $\eta$ which will run over the cube roots of unity, and modify the Boltzmann weight for the lattice gauge theory {on the faces crossed by} $\mathcal{P}$ by the substitution
\be
\notag
\bigg(   1 +  \prod_{l\in F} U_l  + \prod_{l\in F} U_l^*  \bigg)
\longrightarrow
\bigg(  1 + \eta  \prod_{l\in F} U_l  + \eta^* \prod_{l\in F} U_l^*   \bigg).
\ee
The factor $\eta$ goes with the term in which the plaquette's normal is parallel to the path $\mathcal{P}$, and the factor $\eta^*$ with the term where the normal and $\mathcal{P}$ are antiparallel. 

For a given term in the graphical expansion, we can read off the colour of $\mathcal{B}$ from the power of $\eta$ mod three: if all the $\eta^*$s and $\eta$s cancel, $\mathcal{B}$ is red, and if there is an $\eta^*$ or an $\eta$ left over, $\mathcal{B}$ is green or blue respectively. Thus to change the probabilities away from $1/3$ for this cell, we make $\eta$ a dynamical variable and include in the partition function a sum over $\eta$ with the weight
\be
\label{eta sum}
\sum_{\eta = 1, \omega, \omega^2} \lf p_\text{R} + p_\text{G} \eta + p_\text{B} \eta^*  \ri (\dots),
\ee
where the ellipsis stands for the other factors in the partition function. The graphical expansion goes through as before, with each tricolour percolation configuration acquiring a power of $\eta$ determined by the colour of $\mathcal{B}$. Performing the sum over $\eta$, only one of the three terms $p_R$, $p_\text{G} \eta$ or $p_\text{B} \eta^*$ gives a nonzero contribution, yielding the appropriate probability.

We now specialize to the line $p_\text{R}\geq p_\text{G} = p_\text{B}$ (we will briefly discuss the general case in Sec. \ref{vortex derivation}). Writing $\eta = \omega^{\rho_\mathcal{B}}$, where $\rho_\mathcal{B} \in \{0, \pm 1\}$ will become the monopole charge in the cell, the above weight is
\ba
\sum_{\rho_\mathcal{B} = 0, \pm 1} & h^{|\rho_\mathcal{B}|}(\ldots), &  h& = \f{3}{2} \lf p_\text{R} - \f{1}{3} \ri.
\end{align}
At the symmetric point, $\rho_\mathcal{B}$ is forced to be zero, and the modification disappears. Moving away from the symmetric point allows $\rho_\mathcal{B}$ to fluctuate. 

To see the effect of this on the gauge theory, consider the Boltzmann weight in terms of gauge field configurations, rather than in terms of the graphical expansion. The Boltzmann weight for the plaquettes on $\mathcal{P}$ is now maximized not by a magnetic flux $B_F  = 0$ $(\text{mod } 6\pi)$, but by a flux $B_F = 2 \pi \rho_\mathcal{B}$ $(\text{mod } 6\pi)$ oriented antiparallel to $\mathcal{P}$. Thus $\mathcal{P}$ is a Dirac string for a monopole of charge $\rho_\mathcal{B}$ located at $\mathcal{B}$, and $h$ is a fugacity for monopoles.

We may immediately extend the above argument to a uniform (rather than local) change in the probabilities. Each cell acquires a monopole charge $\rho_\mathcal{B}$, and we write
\ba
\notag
Z
=
\sum_{\{\rho_{\mathcal{B}}\}} 
h^{N_\rho}
& \Tr  \prod_{F}  \lf   1 + e^{i(B_F - D_F)/3} + \mathrm{c.c.} \ri \\
 &\phantom{.....} \times\prod_{\<ij\>} \lf 1 +  U_{ij}^3 \psib^\dag_i \psib^{\phantom{\dag}}_j
+  U_{ji}^3 \psib^\dag_j \psib^{\phantom{\dag}}_i \ri.
\label{LGT for TP general}
\end{align}
Here $N_\rho$ is the total number of monopoles, $N_\rho =\sum_{\mathcal{B}} |\rho_\mathcal{B}|$. $B_F$ is the magnetic flux through the face $F$, defined in (\ref{tricolour perc magnetic field}), and $D_F$ is the flux through $F$ due to the Dirac strings $\mathcal{P}$ emanating from the monopoles.   Since the geometry of these strings is arbitrary, $D_F$ is arbitrary except that it must be a multiple of $2\pi$ on every face, and (viewed as a vector field on the dual lattice) must have the right divergence, $(\nabla.D)_{\mathcal{B}}=- 2\pi \rho_\mathcal{B}$. We fix it by any suitable  convention, for example by setting all the Dirac strings to be parallel to a given axis.

A peculiarity of the above mapping is that monopoles appear in two distinct ways: as a result of the microscopic compactness of $U$ and as a result of a modification to the Boltzmann weight for the gauge field. In the continuum derivation (Sec. \ref{vortex derivation}) all monopoles appear in the same way.

\subsubsection{Interpretation}
\label{Interpretation}

\noindent
The lattice model (\ref{LGT for TP general}) describes a supersymmetric matter field,  whose worldlines are the tricords, coupled to a gauge field. The partition function includes a sum over monopole configurations, with monopoles weighted by a fugacity that tends to zero at the symmetric point. If we write a continuum Lagrangian with a gauge field,
\be
\label{TP Lagrangian with B}
\mathcal{L} =|(\nabla- i A)\psib|^2 + \mu |\psib|^2 + \lambda |\psib|^4 + \kappa B^2,
\ee
we must take into account that in the kinetic term $B^2$ for the gauge field we should not include contributions from Dirac strings, and that the monopole fugacity is hidden in the measure for the path integral \cite{Polyakov later paper, Kleinert compact gauge fields} (Sec.~\ref{vortex derivation}). But as mentioned in Sec. \ref{overview}, the universal behaviour is equally well captured by a sigma model formulation. The Lagrangian for this sigma model is given in  Eq.~(\ref{sigma model lagrangian}).

The microscopic form of the action (\ref{LGT for TP general}) does not of course give us the values of the parameters (such as $\mu$ or $g$) in the coarse-grained action. However it is easy to identify the phases of tricolour percolation with the phases of the above theory.

The \emph{extended phase} is the ordered phase of the sigma model at small $g$, or the Higgs phase (at negative renormalized $\mu$) in the language of the gauge theory. The gauge symmetry (Higgs mechanism) removes one Goldstone mode, but $2k$ bosonic and $2k$ fermionic Goldstone modes remain. They are governed by free field theory, and this leads to Brownian behaviour for the vortex lines (see Sec. \ref{statistics of vortex lines} for further details). Note that that while the symmetric point may seem special, in that charge-one monopoles are forbidden there, it is no different from the rest of the extended phase as far as universal properties are concerned.

The \emph{localized phase} is the disordered phase of the sigma model. In the gauge theory, both $\psib$ and $A$ are massive, the latter because of the proliferation of monopoles.

Note that when $k=0$ the two phases collapse into one, and the Goldstone modes which distinguish the extended from the localized phase disappear. This reflects the fact that when $k=0$ we can no longer construct the geometrical correlation functions needed to probe the transition.

\subsection{Vortices in the continuum}
\label{vortex derivation}

\noindent
Focussing on tricolour percolation allowed us to introduce the dual degrees of freedom in an explicit way, but it had the disadvantage that the more {general} applicability of the $CP^{k|k}$ model was obscured by  microscopic details specific to that model, such as threefold anisotropy. In order to make the universality of the result plausible, we now give an alternative derivation in the continuum, building on standard duality arguments for the XY model \cite{banks myerson kogut, thomas, peskin, dasgupta halperin, fisher XY duality, kleinert XY duality, zee}. 

Let $\theta(x)$ denote the phase of the random complex function,
\be
\theta(x)= \arg w(x).
\ee
This field is compact, $\theta(x) \equiv \theta(x)+2\pi$, and ill-defined at the vortices. We expect universal features of their statistics to be independent of the precise details of its probability distribution, so long as correlations in $\theta(x)$ are short-range (just as universal quantities in percolation are robust to changes in the probability distribution of the microscopic variables). Thus we choose the most convenient distribution, taking $\theta(x)$ to be governed by the Lagrangian $\mathcal{L}_\text{XY}$ of the classical XY model,
\be
\label{LXY}
\mathcal{L}_{\text{XY}}=\f{K}{2}(\nabla\theta)^2- H \cos \theta.
\ee
When $H=0$, this has a phase transition at a critical coupling $K_c$, with short-range correlations in the disordered regime $K < K_\text{c}$. The magnetic field $H$ allows us to introduce bias, i.e. to tune the mean value of the random function $w(x)$:  at $K<K_c$ and sufficiently small $|H|$ vortices are in the extended phase, and we enter the localized phase by increasing $|H|$. This gives a critical line separating extended from localized vortices. Note that in this formulation the core energy of a vortex is hidden in the short-distance regularization.

As is well known, the phase transition of the 3D XY model at zero magnetic field is dual to a Higgs transition for a Lagrangian with a \emph{non}-compact gauge field coupled to a single complex scalar $z$ \cite{banks myerson kogut, thomas, peskin, dasgupta halperin, fisher XY duality, kleinert XY duality, zee},
\ba
\label{standard XY duality}
\mathcal{L}_{\text{dual}}= |(\nabla - i A) z|^2 + \kappa (\nabla\times A)^2 + \mu |z|^2 +\lambda |z|^4,
\notag
\end{align}
where the role of the gauge field $A$ is to encode the long-range interactions between vortices which come from integrating over all $\theta(x)$ compatible with a given vortex configuration. The disordered phase of the XY model, in which vortices proliferate, corresponds to the Higgs phase of the gauge theory (at negative renormalized $\mu$), where the field $z$ has condensed and $A$ is massive by the Higgs mechanism. The ordered phase of the XY model corresponds to the disordered phase of the noncompact dual theory in which $z$ is massive and $A$ massless.

This dual theory differs from the one we require in two ways. Firstly, we must consider the theory at nonzero $H$. The operators $e^{\pm i \theta}$ correspond in the dual language to the insertion of magnetic monopoles, so the perturbation $H \cos\theta$ leads to compactness of the gauge field, a point also stressed by Kleinert \cite{kleinert XY duality}. Usually this perturbation would lead simply to a massive theory.  However, in order to capture geometrical correlation functions, we must extend the theory, using supersymmetry to introduce extra degrees of freedom. This allows us to obtain a nontrivial field theory (the $CP^{k|k}$ model) at finite $H$.

Following the general approach described in \cite{zee}, the functional integral over $\theta(x)$ can be split into a `sum' over vortex configurations $\mathcal{C}$, and for each $\mathcal{C}$, an integral over compatible field configurations. To do this, we associate a representative field configuration $\phi(x)$ with each vortex configuration (by any suitable convention), and write $\theta(x)$ in terms of $\phi(x)$ and a vortex-free field $\tilde\theta(x)$:
\be
\label{splitting}
\theta(x) = \phi(x) + \tilde \theta (x).
\ee
$\tilde\theta(x)$ should be treated as a single-valued field in $[-\infty, \infty)$: while Eq.~(\ref{splitting}) defines $\tilde \theta(x)$ only modulo $2\pi$, this ambiguity can be removed using the line integral of $\nabla [\theta-\phi]$.

The path integral over $\theta$ now becomes $\int_\mathcal{C} \int \mathcal{D} \tilde \theta$, where $\int_\mathcal{C}$ is a shorthand for the sum over vortex configurations. To make this precise we would have to fix a regularization of the theory. This will not be necessary for our present purposes; however it is important to note that $\int_\mathcal{C}$ includes a \emph{local} weight determined by physics on the scale of the UV cutoff. This will include an action cost per unit length of vortex, and local interactions between vortex strands. Note that one way to think of the sigma model action (\ref{LXY}) is as an extreme case in which fluctuations in the amplitude of $w\sim e^{i\theta}$ are restricted to to vortex tubes with a thickness on the scale of the short-distance cutoff.

Inserting the expression (\ref{splitting}) for $\theta$ into the action, and making a Hubbard-Stratonovich transformation with a three-component vector field $\xi$ \cite{zee}, one obtains the Lagrangian
\be
\label{LXY2}
\mathcal{L}'_\text{XY} = (2K)^{-1}\xi^2 - i\xi. (\nabla \tilde\theta+ \nabla \phi) - H \cos \theta.
\ee
When $H=0$,  the integral over $\tilde\theta$ forces $\xi$ to be divergence-free. A nonzero $H$ relaxes this constraint, allowing point-like monopole defects where $\nabla.\xi\neq 0$. This can be seen by a standard trick \cite{Polyakov later paper} -- expanding the Boltzmann weight in $H$ gives
\be
\label{fugacity trick}
\exp\lf { H \int \dd^3 x \cos\theta(x) } \ri = \int_{\rho} \exp \lf { i \int \dd^3 x \, \rho(x) \theta(x)} \ri,
\ee
where $\rho(x)$ is a density for a variable number $N$ of monopoles, of strengths $\rho_i=\pm 1$, at locations $r_i$,
\be
\notag
\rho(x)=\sum_{i=1}^N \rho_i \delta(x-r_i),
\ee
and $\int_{\rho}$ is a shorthand for an integral over monopole configurations, weighted by a fugacity $H/2$ per monopole and with a factor of $1/N!$ for indistinguishability:
\be
\label{monopole sum}
\int_\rho\quad
 \text{denotes}\quad
\sum_{N=0}^{\infty} \f{(H/2)^N}{N!} \prod_{i=1}^N \lf \sum_{\rho_i= \pm 1} \int \dd^3 r_i \ri . 
\ee 
Integrating over $\tilde \theta$, the XY partition function is
\ba
\notag
Z_\text{XY}=& 
\int_\mathcal{C} 
\int_\rho
\int \DD\xi \,
\delta(\nabla.\xi-\rho) \\
&\times\exp  \int \dd^3 x \lf -
\f{1}{2K} \xi^2 + i\xi. \nabla\phi + i\rho\phi
\ri.
\label{ZXY with xi}
\end{align}
Since the point-like charges in $\rho$ are quantized, the $2\pi$ ambiguity in $\phi$ does not affect the Boltzmann weight.

We resolve $\xi$ into a divergence-free part $\nabla\times A$ and a collection of Dirac strings $D$ which carry the flux away from the monopoles along singular flux lines, by writing
\ba
\label{split xi}
 \xi &=\f{1}{2\pi}\lf \nabla\times A - D\ri,& &\text{with} & \nabla.D=-2\pi \rho.
\end{align} 
(The geometry of the Dirac strings is chosen by some convention and does not fluctuate.)

As a consequence of (\ref{split xi}), the functional integral over $A$ includes Dirac monopole configurations with singular strings $D$. The form of the kinetic term for the gauge field, involving $\xi^2$ rather than $(\nabla\times A)^2$,  means that these strings do not cost any energy. Actions of this form for compact gauge fields are discussed in \cite{Polyakov later paper, Kleinert compact gauge fields}. 

We now simplify the action in Eq.~(\ref{ZXY with xi}). Since $\nabla \phi$ is not the derivative of a continuous, single-valued function (in terms of such a derivative, $\nabla\phi = - i e^{-i\phi} \nabla e^{i\phi}$), the last two terms do not cancel on integrating by parts. Instead they yield:
\ba
\notag
i \int & \dd^3 x  \lf  \xi. \nabla\phi + \rho\phi \ri  \\ 
=&\f{i}{2\pi} \int \dd^3 x\,A. \lf\nabla\times \nabla\phi\ri+  i \int_{\substack{\text{surfaces of} \\ \text{discontinuity}}} D. \dd^2 S.
\end{align}
The second integral counts intersections of the Dirac strings with the surfaces on which $\phi$ jumps by $2\pi$. Since this is an integer multiplied by $2\pi$, the term does not contribute to the Boltzmann weight. However, the first term provides a coupling of the gauge field to a singular current
\be
J =\f{1}{2\pi} \nabla\times \nabla\phi
\ee
which runs along the vortices. 

At this point we have a theory of fluctuating vortex lines -- flux lines of $J$ -- coupled to a compact gauge field, with partition function
\be
\notag
Z_\text{XY}= \int_\mathcal{C} \int_\rho \int \DD A \exp \left(  - \int \dd^3 x \lf 
\f{1}{2K} \xi^2 + i A. J \ri \right).
\ee
Note that because of the quantization of the monopole charge and of the vortex `flux', the Dirac strings do not couple to $J$. The dependence on $H$ is in the integral  over monopole configurations (\ref{monopole sum}) -- when $H=0$, monopoles are absent and $A$ is noncompact.
\ret
The integral over vortex configurations, so far denoted only symbolically by $\int_\mathcal{C}$, is over configurations of oriented loops, with \emph{local} interactions between segments of loop. The key point of XY duality is that we can equally well view these loops as worldlines of charged particles \cite{zee}. In order to be able to construct geometrical correlation functions, we must introduce additional species of bosonic and fermionic particles, using a supervector $\psib$ rather than a single complex scalar $z$ \cite{Parisi Sourlas SUSY for loops}.  A concrete example of this correspondence between loops and supervectors was displayed in Sec.~\ref{tricolour percolation section}. The $A.J$ term in the action implies that the gauge field couples to $\psib$ in the usual way. The interactions between vortex segments in $\int_\mathcal{C}$ translate into local interactions for $\psib$:
\be
\label{L-dual}
\mathcal{L}_{\text{dual}} = \left|(\nabla-iA)\psib\right|^2 + \mu |\psib|^2 + \lambda |\psib|^4+...
\ee
The line of transitions out of the extended phase at $K<K_c$ and $|H|>0$ is governed by the (compact) $CP^{k|k}$ model. When $H=0$ we obtain instead the $NCCP^{k|k}$ model, a supersymmetric extension of the conventional dual theory for the XY transition. We will consider the phase diagram of this noncompact theory in Sec.~\ref{XY critical point}.

The discussion of tricolour percolation in Sec.~\ref{tricolour percolation section} focussed on the line $p_\text{G}=p_\text{B}$ in the two-dimensional parameter space of the model. As a caricature of tricolour percolation away from this line, we can modify the potential of Eq.~(\ref{LXY}) to $H \cos \theta  + H_3\cos 3 (\theta - \Theta)$. Expanding in $H_3$ leads to the charge three monopoles we expect from Sec.~\ref{tricolour percolation section} (the connection between monopoles and anisotropy has been noted in \cite{kleinert XY duality}). For general $\Theta$ these have a complex fugacity $H_3 e^{\pm 3 i\Theta}/2$, just as Eq.~(\ref{eta sum}) leads to complex monopole fugacities for general $p_\text{R}$, $p_\text{G}$, $p_\text{B}$ \footnote{It is a matter of convention whether the charge one or the charge three monopoles have the non-positive fugacity.}. Changing the bare values of $H_3$ or $\Theta$ is not expected to change critical behaviour in the extended phase or at the $CP^{k|k}$ critical point, where the low energy degrees of freedom are the neutral ones in the supermatrix $Q$. In particular, we expect to see the same behaviour for tricolour percolation everywhere on the critical line \footnote{While a complex fugacity for monopoles may seem  a drastic perturbation, note that this fugacity will naturally renormalize towards positive values. Crudely, after a coarse graining step we cannot distinguish a charge-three monopole from three charge-one monopoles, so the renormalized fugacity $\sim a H_3 e^{3 i\Theta} + b H^3 +\ldots$ with $a, b>0$.}.

\subsection{Statistics of vortex lines}
\label{statistics of vortex lines}

\noindent
Having identified the field theory for vortex lines in short-range correlated $w(x)$, we now use it to answer some natural questions about their statistics.

\subsubsection{Geometrical correlation functions}

\noindent
The geometrical correlation functions of interest to us, the probabilities of various geometrically-defined events, can be expressed as correlation functions of $\psib$s. At a formal level the relationships follow from (\ref{tricolour perc correlator formula}), as indicated in Section \ref{Symmetric point}, but they can easily be anticipated by thinking of vortices as particle worldlines. 

Let $G(x-y)$ be the probability that a single vortex visits both $x$ and $y$. (Strictly, since we are working in the continuum and taking the vortices to be of infinitesimal thickness, we should define a small $\epsilon$ and say `visits regions of size $\epsilon$ around $x$ and $y$'.) At the critical point and in the localized phase all vortices are finite, and the contributions to $G(x-y)$ come from configurations in which a finite loop visits the two points. In the extended phase there are also contributions of a different type, with $x$ and $y$ lying on a single infinite strand. We first discuss the cases where these infinite strands are absent.

The operators needed to construct geometrical correlators are combinations of the gauge-invariant bilinears $Q^{\alpha\beta} = \psi^\alpha \psi^{*\beta} - (\psib^\dag \psib)\delta^{\alpha\beta}$. In fact we will need only $Q^{\alpha\beta}$ for $\alpha,\beta\leq k+1$ and $\alpha\neq\beta$, i.e. $z^\alpha z^{*\beta}$ for $\alpha\neq\beta$. As discussed in Sec.~\ref{overview}, such an insertion restricts the loop expansion of the partition function to configurations in which the point $x$ has an incoming strand of colour $\alpha$ and an outgoing strand of colour $\beta$ -- or equivalently, configurations in which there is a loop passing through $x$ which changes colour from $\alpha$ to $\beta$ at $x$. Thus we see that the correlation function $\<Q^{\alpha\beta}(x) Q^{\beta \alpha}(y)\>$ forces a single loop to visit both $x$ and $y$:
\be
G(x-y) \propto \<Q^{12}(x) Q^{21}(y) \>.
\ee
Similarly, the probability that $m$ loops connect the vicinity of $x$ to the vicinity of $y$ scales like $\< Q^{12}(x)^m Q^{21}(y)^m \>$. As a further example, the probability that the three sites $x$, $y$ and $z$ all lie on the same vortex can be written $F(x,y,z)\propto\<Q^{12}(x) Q^{23}(y) Q^{31}(z) \>$.

\subsubsection{Vortices in the critical region}
\label{critical loops}

\noindent
At the critical point, the two-point function decays as $G(x)\sim 1/|x|^{1+\eta}$, where $\eta$ is the anomalous dimension of $Q$. Moving slightly away from criticality, a finite correlation length appears, scaling as $\xi\sim |h-h_c|^{-\nu}$ if $h$ parametrizes the bias: in the localized (disordered) phase $\xi$ is a measure of the typical size of a large loop (e.g. $G(x)$ decays with the Ornstein-Zernicke form $|x|^{-1}e^{-|x|/\xi}$), and in the extended (ordered) phase, anticipating the next subsection, it sets the scale beyond which correlations are Brownian. In this phase $Q$ aquires a nonzero expectation value $\<Q\>\sim (h_c-h)^{\beta}$, which gives the probability that a given point lies within a fixed small distance of an infinite vortex, or that a given link in tricolour percolation lies on an infinite tricord.

A scaling argument \cite{duplantier saleur percolation, scaling relations} relates $\eta$ to the fractal dimension $d_f$ of the critical vortices, and to the exponent $\tau$ governing the number density $n(l)$ of vortex loops of length $l$. With $n(l)\sim l^{-\tau}$,
\ba
\label{df tau scaling rels}
d_f& = \f{5-\eta}{2} & \tau= \f{11-\eta}{5-\eta}.
\end{align}
Hyperscaling in 3D giving $\beta = \nu (1+\eta)/2$.
 
Precise numerical values are available from simulations of a lattice loop model \cite{ortuno somoza chalker} which is also in the $CP^{k|k}$ universality class \cite{short loop paper}:
\ba
\label{accurate critical exponents}
d_f &= 2.534(9), & \tau& = 2.184(3), & \nu&= 0.9985 (15).
\end{align}
This yields $\eta = -0.068(18)$.

Clearly this critical point is not accurately described by mean field theory, which would require $\nu = 1/2$. However it is interesting to consider what the mean field prediction for $d_f$ is: surprisingly, it is not the `trivial' value $d_f^\text{MF}=2$, but the fairly accurate value $d_f^\text{MF}=5/2$. The reason one might have expected $d_f^\text{MF}=2$ is that mean field theory corresponds to a free field description of the critical point, and free field theory is usually associated with Brownian walks (as in the extended phase, where free field theory for the Goldstone modes leads to Brownian behaviour). Here, however, that is not the case, because the appropriate free theory is formulated in terms not of $\psib$ but of the composite field $Q$: in a replica approach,
\be
\mathcal{L}_\text{free} = \tr\, (\nabla Q)^2.
\ee
This leads to $\eta=0$ and a fractal dimension $d_f^\text{MF} = 5/2$. Since $\eta$ is small, as is typical for critical points in 3D, this approximation is quite accurate. The fact that $d_f$ is close to $5/2$ rather than to two is a clear symptom of confinement and the fact that vortices are worldlines not of $Q$, but of its `square root', $\psib$. (In the context of quantum magnets described by $CP^1$/$NCCP^1$ models \cite{NCCPn-1}, a \emph{large} value of $\eta$ has been used as a diagnostic for deconfinement, i.e. the non-compactness of the $U(1)$ gauge field.) The above assumes a soft spin formulation for $Q$, which we discuss further in Sec.~\ref{Other universality classes}.

\subsubsection{Extended phase: Goldstone modes and Brownian loops}
\label{extended phase}

\begin{figure}[t] 
\centering
\includegraphics[width=1.7in]{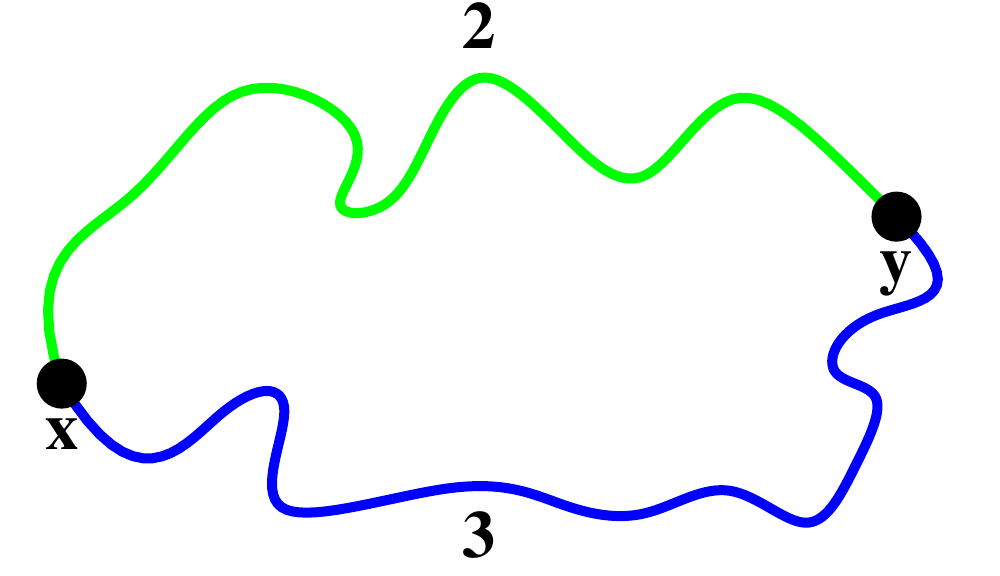}\includegraphics[width=1.7in]{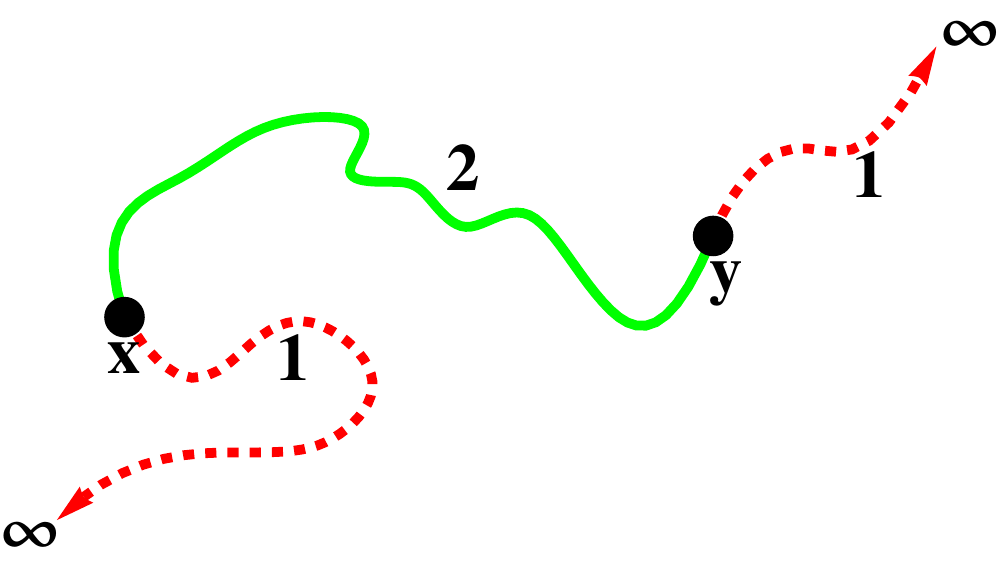}
\caption{(Color online) Dominant contributions to the correlation functions $\<Q^{23}(x)Q^{32}(y)\>\sim |x-y|^{-2}$ and $\<Q^{12}(x)Q^{21}(y)\>\sim |x-y|^{-1}$ in the extended phase as described in the text.  The orientations of the loops are not indicated.}
\label{QQ correlator fig}
\end{figure}

\noindent
In three dimensions, a random walker need never return to his starting point, but can escape to infinity. The same is true of vortices in the extended phase. Of course, in a finite system there is no such thing as escape to infinity: what takes its place depends on the boundary conditions. The simplest and most natural choice is to allow vortices to terminate on the boundary. The possibility of escaping to infinity then means that a vortex strand passing through the origin has a finite probability of reaching the boundary, even in the thermodynamic limit. If the linear size of the system is $L$, then these system-spanning lines have an arc length of order $L^2$, since their fractal dimension is two. On the other hand, if we were to impose periodic boundary conditions, these lines would join up into loops with typical length $L^3$ \cite{Austin, Jaubert, short loop paper}.

In order to discuss geometric correlation functions in the extended phase we must separate out the `infinite' strands.  We take a finite system in which vortex lines can end on the boundary, and require that any line which does so has a specific (bosonic) colour, say $\alpha = 1$. With this protocol the fugacity per vortex is still one, regardless of whether the vortex forms a closed loop or ends on the boundary. In the $CP^{k|k}$ language, these boundary conditions correspond to applying a symmetry-breaking field for the $CP^{k|k}$ order parameter on the boundary \footnote{For tricolour percolation this is done by modifying the Boltzmann weight for the boundary links, replacing $U_{ij}^3\psib_i^\dag \psib_j$ with $U_{ij}^3 z^{*1}_i z_j^1$. In the scaling limit this yields a boundary magnetic field having the effect described.}. The extended phase is the ordered phase of the field theory, and this boundary condition fixes the direction of the order parameter to be $\psib \propto (1,0,...,0)$. We may fix the gauge freedom by taking the first component real and positive. Fluctuations involve $k$ (complex) bosonic Goldstone modes, $\phi^2$,...,$\phi^{k+1}$, as well as $k$ fermionic ones, which are just the original fermions $\chi$: $\psib \sim (\sqrt{1- |\phi|^2 -|\chi|^2}, \phi, \chi)$. To leading order in these modes (for $\alpha,\beta=2,...,k+1$),
\ba
\notag
Q^{11} & \sim 1, &
Q^{\alpha 1} & \sim \phi^\alpha, &
Q^{\alpha \beta} & \sim \phi^\alpha \phi^{*\beta}.
\end{align}
Since $\< \phi^{\alpha} (x) \phi^{*\beta}(y)\> \propto \delta^{\alpha\beta} |x-y|^{-1}$ we therefore have
\ba
\notag
\< Q^{23}(x)  Q^{32} (y) \> & \sim \f{1}{|x-y|^2},
 & \< Q^{12}(x)  Q^{21} (y) \> & \sim \f{1}{|x-y|}.
\end{align}
The first of these gives the probability that $x$ and $y$ lie on the same \emph{finite} loop, since strands of colours 2 and 3 are prohibited from escaping to infinity. On the other hand, the second of them is dominated by configurations in which a single \emph{infinite} vortex passes through both points -- that is, they are connected by a single finite strand, rather than by two strands as for points sharing a finite vortex.

In general, one finds that the probability $G_M(x)$ that two regions separated by a distance $|x|$ are connected by $M$ (finite) strands of vortex scales like $|x|^{-M}$:
\be
\notag
G_{M}(x) \sim |x|^{-M}.
\ee
(The correlator $G$ discussed in the context of the localized phase and critical point corresponds to $G_2$; in this regime $G_M=0$ for odd $M$.) These simple exponents indicate that the vortices behave on large scales like free random walks -- each factor of $1/|x|$ is just the probability that a random walker, starting in one of the regions, happens to visit the other. It follows that the fractal dimension is two, and the number density of loops of length $l$ behaves as $n(l) \sim l^{-5/2}$. 

As another example, we easily check that the probability that three distant sites lie on the same vortex behaves as
\ba
\notag
F(x_1,x_2,x_3) &\sim \f{r_{12}+r_{23}+r_{31}}{r_{12}r_{23}r_{31}}, & r_{ij}&=|x_i-x_j|,
\end{align}
being dominated by contributions from infinite vortices, while the probability that they lie on the same finite vortex scales as $(r_{12}r_{23}r_{31})^{-1}$. 

This result -- Brownian statistics for vortices in the extended phase -- is expected on the basis of numerical simulations \cite{Vachaspati Vilenkin, Scherrer Frieman, Strobl Hindmarsh}. Surprisingly, a clear theoretical derivation has not previously been provided, although a heuristic explanation \cite{Scherrer Frieman} has been given by  analogy with polymers in the melt \cite{polymers in melt}.

\section{Variations and perturbations}
\label{Other universality classes}

\noindent
So far, we have discussed universal behaviour for vortices in short-range correlated environments. In this section we consider the stability of this universality class to perturbations, including long-range correlations in the complex field and a perturbation leading to a crossover to conventional percolation, and also the universal behaviour of other kinds of vortex problem: unoriented vortices in nematics, and vortices in the vicinity of the XY critical point. We also briefly discuss the $6-\epsilon$ expansion for vortices.

A useful language for considering perturbations of the $CP^{k|k}$ model is the soft spin Lagrangian for the gauge-invariant composite field $Q$, Eq.~(\ref{first appearance of soft spin theory}). Since some perturbations cannot be written in the supersymmetric theory, and also for simplicity, we use the replica formulation in which $Q$ is an $n\times n$ traceless Hermitian matrix with the implied limit $n\rightarrow 1$. 
 
Microscopically, this matrix is given by $Q = \zb \zb^\dag - 1/n$ with $\zb^\dag \zb =1$, so satisfies the nonlinear constraint $(Q+1/n)^2 = (Q+1/n)$ in addition to the linear one $\tr Q =0$. In a soft spin approach, we might imagine imposing a softened version of the former via a potential 
\ba
\notag
\tr &\lf (Q+1/n)^2 - (Q+1/n) \ri^2 \\ &= \text{const.} + \f{n^2-6n+6}{n^2} \tr Q^2 + \f{2(2-n)}{n} \tr Q^3 + \tr Q^4.
\notag
\end{align}
Of course these `bare' coefficient values have no significance, except to draw our attention to the fact that (except in the case $n=2$, where there is an additional symmetry $Q\rightarrow -Q$) there is a symmetry-allowed cubic term \cite{duane green, McKane Wallace Zia, Senthil UCD}. For $n>2$, this term implies that the mean field prediction (expected to be valid in a large number of spatial dimensions) is a first order transition. However, this conclusion does not apply to the replica limit $n\rightarrow 1$. Such a situation is familiar from percolation \cite{6-epsilon for percolation}.

\subsection{Upper critical dimension and $6-\epsilon$ expansion}
\label{upper critical dimension}
 
 \noindent
The cubic term implies that the upper critical dimension for $CP^{n-1}$ in the limit $n\rightarrow 1$ (and $CP^{k|k}$) is expected to be six. Certain correlation functions for Anderson localization in symmetry class C are also described by this field theory, and can in fact be mapped to correlation functions in classical loop models described by $CP^{k|k}$ \cite{glr, mirlin evers review, beamond cardy chalker, ortuno somoza chalker, read saleur, short loop paper}, so we expect six to be an upper critical dimension for that problem too \cite{Senthil UCD}.

We may also consider the $6-\epsilon$ expansion, which was performed for theories with a single cubic invariant in \cite{Amit, McKane Wallace Zia, Bonfim}. Eq.~(\ref{first appearance of soft spin theory}) at $n>1$ is precisely the theory considered in \cite{McKane Wallace Zia}, in the form given by writing $Q$ in terms of $(n^2-1)$ real scalar fields $t_i$ and the $SU(n)$ generators $T_i$ via $Q =t_i T_i$. With the tensor $d_{ijk}$ defined by $\{T_i, T_j\} = \f{1}{n} \delta_{ij} + d_{ijk} T_k$,
\be
\mathcal{L}_\text{soft spin} = \f{1}{2} (\nabla t)^2 + \f{\mu}{2} t^2 + \f{\nu}{4} d_{ijk} t_i t_j t_k.
\ee
Coefficients in the $\epsilon$ expansion are given by contractions of the $d$s \cite{Amit}. In the  limit $n\rightarrow 1$, to order $\epsilon^2$, 
\ba
\label{epsilon expansion results}
\nu & = \f{1}{2} + \f{5 \epsilon}{76} + \f{733 \epsilon^2}{27436} + \ldots,
&
\eta &= -\f{\epsilon}{19} - \f{166\epsilon^2}{6859} + \ldots.
\end{align}
In three dimensions, this gives $\nu \simeq 0.94$ and $\eta\simeq -0.38$, to be compared with Eq.~(\ref{accurate critical exponents}).  The expressions Eq.~(\ref{epsilon expansion results}) are compatible with the $O(\epsilon)$ results for class C localization in \cite{Senthil UCD}. Note that the sigma model formulation Eq.~(\ref{sigma model lagrangian}) leads to an alternative approximation scheme for exponents via the $2+\epsilon$ expansion, which has been performed to high orders -- see \cite{mirlin evers review} and references therein.

 The field theory in higher dimensions can again be related to loops, and it is natural to conjecture that in $d$ dimensions these have the statistics of zero lines in a random $(d-1)$-component vector field, $(w_1,...,w_{d-1})$. (Such lines are oriented in any number of dimensions, with orientation  given by the vector $\epsilon_{\mu_1...\mu_d} \partial_{\mu_2}w_1...\partial_{\mu_d}w_{d-1}$.)

\subsection{Vortex gluing and percolation}
\label{crossover to percolation}

\noindent
In cases where intersections of vortices cannot be neglected (such as in the lattice XY model), a convention is required for defining geometrical correlation functions, or in other words for determining which vortex strands are regarded as connected. We may either adopt a convention in which vortices remain topologically one-dimensional, or we may `glue' intersecting strands together to form branching, net-like clusters \cite{Kajantjie, Bittner}, and our choice may affect the observed critical behaviour. We now address this issue. The punchline is that with the former convention (when vortices are topologically one-dimensional), the results of the supersymmetric dual theory continue to apply even in the presence of intersections, whereas vortex gluing induces a crossover to conventional percolation. This crossover can be seen in the soft spin language. Another reason for clarifying the relation to percolation is that it has previously been believed that some or all of the exponents for tricolour percolation coincide with those for conventional percolation \cite{tricolour percolation, Strobl Hindmarsh, Schakel entangled vortices}. The soft spin formulation makes clear that this is not the case.

\begin{figure}[h] 
\centering
\includegraphics[width=0.9in]{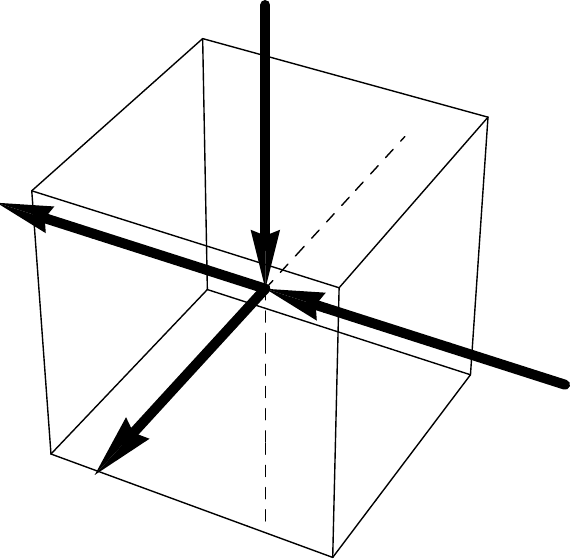}
\includegraphics[width=0.9in]{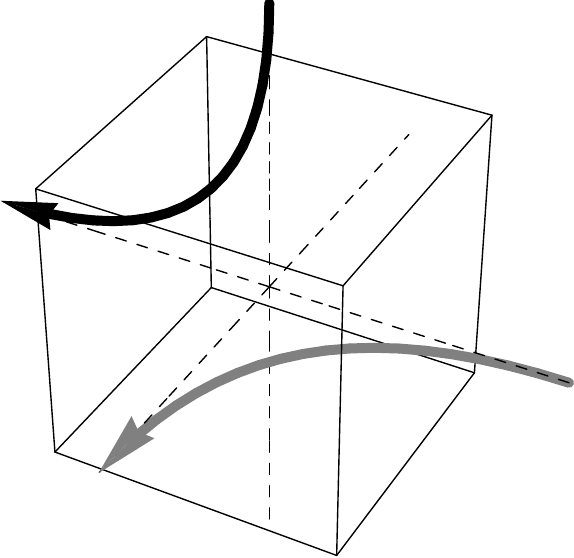}
\includegraphics[width=0.9in]{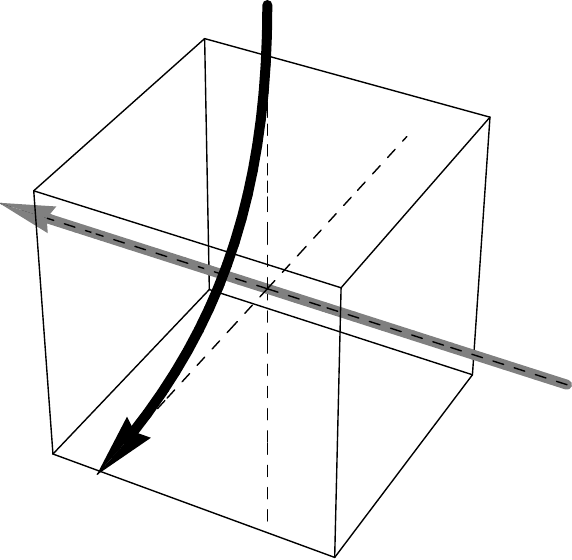}
\caption{Left: vortex strands intersecting at a site of the dual lattice (cube of the original lattice) in a cubic lattice XY model. According to the `stochastic' rule, we choose randomly from the ways of resolving the intersection (centre, right).}
\label{vortex intersection}
\end{figure}

A microscopic model with intersections can be got by modifying the lattice for tricolour percolation (specifically, we can replace the bcc with the fcc lattice, so the cells are rhombic dodecahedra). Alternately, we can use the lattice XY model. Considering such models shows that the dual SUSY theory ($CP^{k|k}$ or $NCCP^{k|k}$) still applies if we resolve vortex intersections according to the `stochastic rule' used in some numerical simulations \cite{Kajantjie, Bittner}: at an intersection with $q$ incoming and $q$ outgoing vortex strands, we choose randomly from the $q!$ ways of pairing the former with the latter (Fig.~\ref{vortex intersection}). This rule is a simple consequence of formulas like Eq.~(\ref{single link traces}) for the traces at a single site.

To induce a crossover to percolation, we modify the rule by gluing intersecting vortices with a probability $c>0$ \cite{Kajantjie, Bittner} \footnote{In terms of connectivity, setting $c=1$ is equivalent to the `maximal' definition of \cite{Kajantjie}.}. In the replica language, each strand acquires a colour $\alpha=1,\dots n$, and to glue strands we must force their colours to coincide; the probability $G_\text{perc}$ of two links lying in the same cluster will then be given by
\be
G_\text{perc} \sim \<Q^{\alpha\alpha}Q^{\alpha\alpha}\>,
\ee
being simply related to the probabilities of their being the same or different colours. We can check in the graphical expansion of a lattice model with intersections (we omit details) that the perturbation induced by the gluing is of the form $\delta \mathcal{L} \sim  - \, c \, \sum_\alpha |z^\alpha|^4$. 
As expected, this operator creates a meeting of four worldline strands of the same colour. The perturbation breaks rotational symmetry in replica space, yielding an additional mass term for the off-diagonal elements of Q in $\mathcal{L}_\text{soft spin}$. Subtracting a multiple of $\tr Q^2$, which in the absence of the symmetry-breaking term would merely shift the critical point,
\be
\delta\widetilde{\mathcal{L}} \sim \sum_{\alpha\neq \beta} |Q^{\alpha\beta}|^2.
\ee
$\delta\widetilde{\mathcal{L}}$ is expected, on the basis of its naive scaling dimension, to be relevant at the $CP^{k|k}$ critical point. (Both $\delta\widetilde{\mathcal{L}}$ and the perturbation discussed in Sec. 4C can be written in terms of the traceless symmetric tensor $S^{\alpha\gamma}_{\beta\delta}$ built out of $Q^{\alpha \beta}Q^{\gamma \delta}$, modulo a $\tr Q^2$ term.) Symmetry considerations then lead us to anticipate a crossover to behaviour in the universality class of the $n$-state Potts model in the limit $n\rightarrow 1$, which is a well-known description of percolation \cite{cardy book}.  Note that the Goldstone modes in the extended phase acquire a mass, consistent with the fact that in conventional percolation the percolating phase is massive.

The above picture implies that the bias-induced vortex transition is in a distinct universality class from that of percolation. Previously this has been obscured by the fact that some exponents are similar \cite{tricolour percolation, Strobl Hindmarsh, Schakel entangled vortices}. In particular, the fractal dimension of critical tricords was found \cite{tricolour percolation} to be $d_f = 2.54(1)$ (see also  Eq.~(\ref{accurate critical exponents})), close to that of critical percolation clusters, for which a more recent study \cite{percolation numerics} gives $d^\text{perc}_f = 2.5226(1)$. In the light of the above, this similarity is not surprising, and just reflects the small anomalous dimension of $Q$ at each of respective critical points. (The result $d^\text{perc}_f\sim 5/2$ can also be seen without invoking field theory, via a Flory-like argument due to de Gennes \cite{de Gennes animals vs percolation}.) The correlation length exponents for the two transitions are clearly different -- compare  $\nu^\text{perc}= 0.8733(5)$ \cite{percolation numerics} with Eq.~(\ref{accurate critical exponents}).

\subsection{Unoriented vortices}
\label{unoriented vortices}

\begin{figure}[h] 
\centering
\includegraphics[width=1.8in]{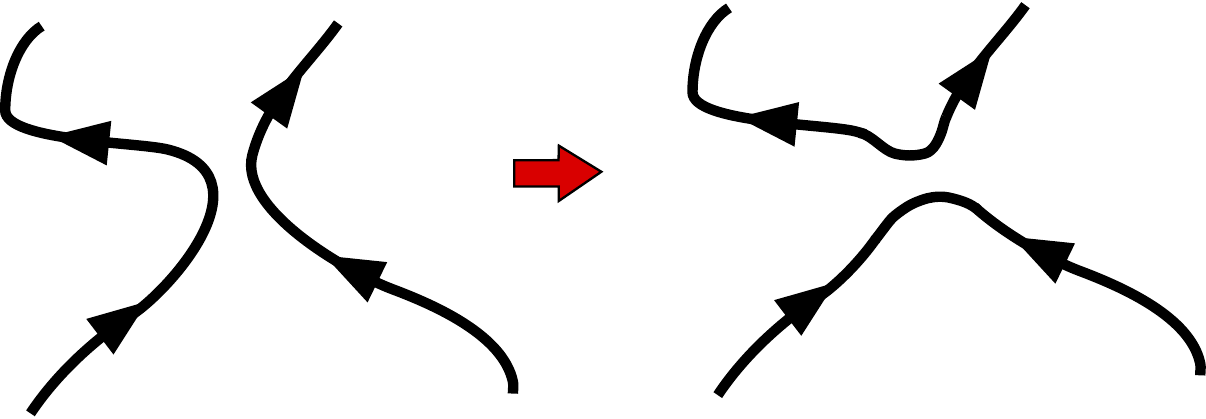}
\caption{(Color online) `Rewiring' of vortices induced by perturbing the oriented vortex problem.}
\label{rewiring}
\end{figure}

\noindent
So far we have discussed oriented vortices in random complex fields, which are associated with the fundamental group $\pi_1(S^1)=\mathbb{Z}$. It is natural to expect unoriented vortices, which appear when the relevant fundamental group is equal to $\mathbb{Z}_2$, to correspond to a different field theory, since the distinction between positively and negatively charged fields ($\psib$ and $\psib^*$) in the $CP^{k|k}$ model is due to the distinction between vortices and antivortices. To obtain the field theory for $\mathbb{Z}_2$ vortices from the $CP^{k|k}$ model, we must introduce a perturbation which reduces the symmetry (from superunitary to orthosymplectic); the introduction of crossings into 2D loop models for non-crossing loops leads to the same symmetry breaking \cite{dense loops and supersymmetry}.

Unoriented vortices occur (for example) in nematic order parameters, where the order parameter manifold is $RP^{N-1}$ with $N>2$. This is the space of $N$-component unit vectors ${\boldsymbol v}$ with the identification ${\boldsymbol v}\sim -{\boldsymbol v}$. Vortices of this kind, for $N=3$, have been observed in nematic liquid crystals \cite{Bowick, Chuang} and their fractal geometry has been studied numerically in the context of cosmic strings \cite{kibble Z2, Strobl Hindmarsh}.

As for oriented vortices, we may consider the phase diagram of vortices in short-range correlated $RP^{N-1}$ fields as a function of bias favouring some point in the order parameter manifold. At small bias, there is an extended phase similar to that for oriented vortices \cite{kibble Z2, Strobl Hindmarsh}. We argue below that this phase is again Brownian (thus we disagree with earlier suggestions that the fractal dimension is less than two in this phase \cite{Strobl Hindmarsh}). However, this Brownian behaviour results from the Goldstone phase of a field theory different to the $CP^{k|k}$ model, so the bias-induced critical point is in a different universality class to that for oriented vortices. The initial investigation of this transition by Strobl and Hindmarsh \cite{Strobl Hindmarsh} is probably not sufficiently precise to confirm this, and further numerical work would be interesting.

To determine the field theory, we access the unoriented vortex problem by applying a perturbation to that of oriented vortices. Take an $RP^2$ order parameter ${\boldsymbol v}=(v_1, v_2,v_3)$ with ${\boldsymbol v}^2=1$ and ${\boldsymbol v}\sim -{\boldsymbol v}$. Now, the subspace $v_3 = 0$ is $RP^1=S^1$; thus in the limit where fluctuations in $v_3$ are completely suppressed, we return to the oriented vortex problem. Turning fluctuations in $v_3$ back on allows the orientation of a vortex to fluctuate along its length. Specifically, the leading effect of reintroducing fluctuations in $v_3$ is to allow configurations of the type shown in Fig.~\ref{rewiring}, where two vortices passing close to each other are rewired in a manner `incompatible' with the orientations of the strands. 

This perturbation is most easily discussed in the replica theory. In terms of wordlines, the right hand side of Fig.~\ref{rewiring} corresponds to the annihilation of two particles of the same colour $\beta$, and the creation of two particles of colour $\alpha$, so the perturbation leads to the following $SU(n)$-symmetry-breaking term:
\be
\label{orientation breaking perturbation}
\delta \mathcal{L} = - \sum_{\alpha,\beta=1}^n (z^{*\alpha} z^{*\alpha})
 (z^{\beta} z^{\beta}) = - |z^T z|^2.
\ee
In the soft-spin theory for $Q$, which we decompose into real and imaginary parts, $Q = Q_R + i Q_I$, this yields
\be
\delta \widetilde{\mathcal{L}} \sim \sum_{\alpha,\beta=1}^n (Q_I^{\alpha\beta})^2.
\ee
We see that allowing fluctuations of the vortex orientation yields a mass term for the imaginary part of $Q$. The simplest assumption is that this leads to a crossover to a similar theory but with real rather than complex $Q$.  The resulting traceless symmetric matrix is the order parameter for $RP^{n-1}$ as opposed to $CP^{n-1}$. (This $RP^{n-1}$ field should not be confused with the $RP^{N-1}$ field which hosts the vortices.) This can be easily checked in the case $n=2$, where the $CP^{n-1}$ model is the $O(3)$ model, and where the perturbation is a mass term for one of the components of the $O(3)$ spin, leading to a crossover to the XY, or equivalently $RP^1$, universality class.

So we expect that unoriented vortices in a (short-range correlated) random nematic field are described by the replica limit of the $RP^{n-1}$ sigma model. Although for brevity we have given only a perturbative argument, additional evidence comes from the existence of lattice loop models with $RP^{n-1}$ degrees of freedom \cite{short loop paper}. It is also possible to construct a tricolour-percolation-like model which can be formulated as a lattice gauge theory with the same symmetry.

Note that in the extended phase it makes no difference whether we think of the field theory for unoriented vortices as the replica limit of an $RP^{n-1}$ sigma model, with target space $RP^{n-1}=S^{n-1}/\mathbb{Z}_2$, or an $O(n)$ sigma model with target space $S^{n-1}$. The topological difference between these manifolds is not seen by the Goldstone modes. However the difference does become important at the critical point: the $O(1)$ sigma model has a {thermodynamic} (Ising) transition, whereas the vortex problem has a geometric transition which is invisible in thermodynamic quantities. This is reflected in the fact that the manifold $RP^{n-1}$ becomes a point when $n=1$.

In SUSY language, the replica limit of the $RP^{n-1}$ model becomes a sigma model for a `real' supervector field $\Phi = (S^1,...,S^{2l+1}, \eta^1,..., \eta^{2l} )$, with $\Phi \Lambda \Phi = 1$, where
\ba
\notag
\Lambda &=
\left(
\begin{array}{ccc}
1_{2l+1}  & 0  & 0  \\
0  & 0  & 1_l  \\
 0 & -1_l & 0  
\end{array}
\right)
\end{align}
(the subscripts give the dimensions of the unit matrices), similar to the 2D supersymmetrized $O(n)$ model described in \cite{dense loops and supersymmetry} in the context of lattice loop models with unoriented loops, but with the additional  identification $\Phi \sim - \Phi$. Note that we can also cast the perturbation (\ref{orientation breaking perturbation}) in supersymmetric language. We must take the $CP^{k|k}$ model with $k=2l$. Then (\ref{orientation breaking perturbation}) should be replaced with
\ba
\notag
\delta\mathcal{L} &= 
- (\psib^T \Lambda \psib) (\psib^\dag \Lambda \psib^*).
\end{align}
It can be checked in a lattice model (such as the modification of tricolour percolation described in \ref{crossover to percolation}) that this is the desired perturbation and that it gives all loops the correct fugacity of one.

\subsection{Stability to long range correlations in $w(x)$}
\label{long range correlations}

\noindent
We have discussed short-range correlated $w(x)$ at length. How short-ranged do the correlations have to be, in order not to alter the universal behaviour in the extended phase or at the critical point? 

Weak correlations in $w(x)$ can be addressed in a standard way using the extended Harris criterion \cite{Weinrib, cardy book}. Correlations in $|w|$ are more relevant than correlations in $\arg w$, and we will consider only the former. Letting these correlations decay as $1/ |x-y|^{A}$ at large distances, one finds that the extended phase is extremely robust, being unaffected by sufficiently weak correlations for any $A>0$. (The statistics of vortices in correlated  complex fields have been studied numerically in \cite{Scherrer Vilenkin}, and the results appear compatible with this claim.) The critical theory on the other hand is stable so long as
\ba
\label{short range condition}
A &> 2/\nu \simeq 2.003(3).
\end{align}
These criteria may be derived by viewing correlations in $w(x)$ as due to quenched, correlated disorder in the parameters of the probability distribution for $w$ \cite{Weinrib} (for example, quenched disorder in the parameters $p_\text{R}$, $p_\text{G}$ and $p_\text{B}$ of tricolour percolation) and   averaging over this disorder to give an effective Lagrangian with non-local couplings between the operators of the $CP^{k|k}$ model, whose relevance or irrelevance can then be determined. This average does not require use of the replica trick, since the partition function takes a trivial value which is independent of the disorder realization.

In addition, correlations decaying \emph{slower} than Eq.~(\ref{short range condition}) demands may still count as short-range if they oscillate with distance \cite{Bogomolny Schmit}. This phenomenon occurs for random superpositions of plane waves $\exp{(i k. x)}$ with fixed $|k|$. According to Berry's random wave model \cite{Berry}, such superpositions
\be
\label{random wave model}
w (x) = \int_{k^2=E} \dd^{d-1} k \,  a(k) e^{i k. x},
\ee
with Gaussian, delta-correlated $a(k)$, give a good description of the statistics of  eigenfunctions of the Laplacian,
\be
-\nabla^2 w = E w,
\ee
in chaotic $d$-dimensional domains. In 2D, the version of this problem with real $w$ has been studied extensively in the context of quantum chaos \cite{nodal lines}, and the nodal lines $w(x)=0$ have been found to have the statistics of percolation cluster boundaries. Bogomolny and Schmit \cite{Bogomolny Schmit} reconciled this with the square root decay of the correlation function at large distances, $\< w(x) w(y) \> \sim \cos(E^{1/2} |x-y|-\pi/4) / \sqrt{|x-y|}$ -- which would usually be slow enough to make percolation results inapplicable \cite{Weinrib} -- by taking oscillations into account. A simple generalization of this analysis shows that in 3D, for complex $w$, the collapse transition of the lines $w(x) = h$ induced by increasing $|h|$ is expected to be in the $CP^{k|k}$ universality class, despite the slow decay of the correlator $\< w^*(x) w(y)\> \propto {\sin(E^{1/2} |x-y|)}/{ |x-y|}$.

\subsection{Vortices near the XY critical point}
\label{XY critical point}

\noindent
It has long been recognized that vortices play an important role in the phase transition of the 3D XY model. XY duality makes this more concrete, giving a dual formulation in which the transition is represented by the condensation of a `vortex' field. A natural question is thus whether the thermodynamic phase transition coincides with the geometric phase transition for the vortices, or whether the two transitions are separate. Simulations of lattice XY models \cite{Kajantjie, Bittner, Hove 2} suggest that the geometric transition occurs slightly inside the XY ordered phase, but very close to the the thermodynamic transition (letting $K$ be the coupling for an XY model on the cubic lattice, the most accurate determination of the geometrical critical point in Ref.~\cite{Kajantjie} gives $({ K_c^\text{geom} - K_c^\text{therm} })/{ K_c^\text{therm}}\sim 4 \times 10^{-4}$). Vortices must of course be defined using the stochastic rule (Sec.~\ref{crossover to percolation}).

At first sight, the $NCCP^{k|k}$ formulation of this problem
\be
\notag
\mathcal{L}_{NCCP^{k|k}} = \left|(\nabla-iA)\psib\right|^2 + \kappa (\nabla\times A)^2 + \mu |\psib|^2 + \lambda |\psib|^4
\ee
 implies that the transitions should coincide, occurring where the field $\psib$ condenses. However, this assumption is not justified, because in the theory with a noncompact gauge field we must distinguish between the condensation of $\psib$ and the condensation of the bilinear $Q$.

The geometric transition into the extended phase is signaled by the appearance of an expectation value of $Q$, i.e. by the spontaneous breaking of supersymmetry: $\<Q\> \neq 0$. On the other hand the thermodynamic transition into the XY disordered phase is signaled by the onset of the Higgs mechanism, due to the condensation of $\psib$, and the generation of a mass for the gauge field. Following standard convention we denote this $\<\psib\>\neq 0$ (though of course this expectation value is zero by gauge invariance). In the simplest scenario, these events would occur simultaneously. However the numerical results \cite{Kajantjie, Bittner} \footnote{While \cite{Kajantjie, Bittner} appear to show two separate transitions, scaling collapse for the geometrical observables in \cite{Bittner} required the use of the standard XY correlation length exponent (A. M. J. Schakel and E. Bittner, private communication), which is surprising unless either it or the splitting of the transitions can be attributed to finite size effects.} lead us to consider the possibility of a phase with $\< \psib\>=0$ but $\<Q \> \neq 0$. Heuristically, this is a pair condensate \cite{Motrunich Vishwanath} of vortices in which the XY order is not disrupted because vortices and antivortices are locally paired.

If the geometric transition does occur within the XY ordered phase in (for example) lattice XY models, a natural guess is that it is in the same universality class as the bias-induced transition. Note that the nonzero value of the XY order parameter yields a bias. In this scenario, long distance behaviour precisely at the geometrical phase transition would be described by a Lagrangian of the form 
\be
\mathcal{L} = \kappa (\nabla\times A)^2 + \mathcal{L}_{\text{critical }CP^{k|k}}[Q],
\ee
where the massless \emph{noncompact} gauge field $A$ (describing Goldstone fluctuations in the XY order) is decoupled, at long wavelengths, from the neutral degrees of freedom in $Q$. These are then described by a critical \emph{compact} $CP^{k|k}$ model. The symmetry-allowed couplings between $A$ and $Q$ are irrelevant at this critical point. This scenario is also compatible with the extended Harris criterion (note that in the Goldstone phase correlations in the modulus of the complex field decay much faster than correlations in its argument).

Previous attempts have been made to relate the fractal dimension of vortices in critical XY or Abelian Higgs models to local correlators in field theory using the duality between these two theories \cite{Hove, Hove 2, Hove reply}. However, the authors did not make use of SUSY or replica, which are necessary in order to form nontrivial geometrical correlators. As a result the scaling relations put forward were not correct, as has been previously noted \cite{Prokofev Svistunov, Janke Schakel}. Note that vortices in Abelian Higgs models, with Lagrangian $\mathcal{L}_\text{SC}= |(\nabla - i a) \Phi |^2 + \tilde\kappa (\nabla\times a)^2 + V(\Phi)$, can be treated in a similar manner to vortices in ungauged fields, leading to a dual SUSY theory without a gauge field (Eq.~\ref{wrong lagrangian}) \footnote{Here $a$ is a dynamical field. The phase structure of the XY model in a fixed external gauge field has been considered in Z.~Te\v{s}anovi\'{c}, Phys. Rev. B {\bf 59}, 6449 (1999).}.

\section{Loops in two dimensions}
\label{percolation}

\begin{figure}[h] 
\centering
\includegraphics[width=2in]{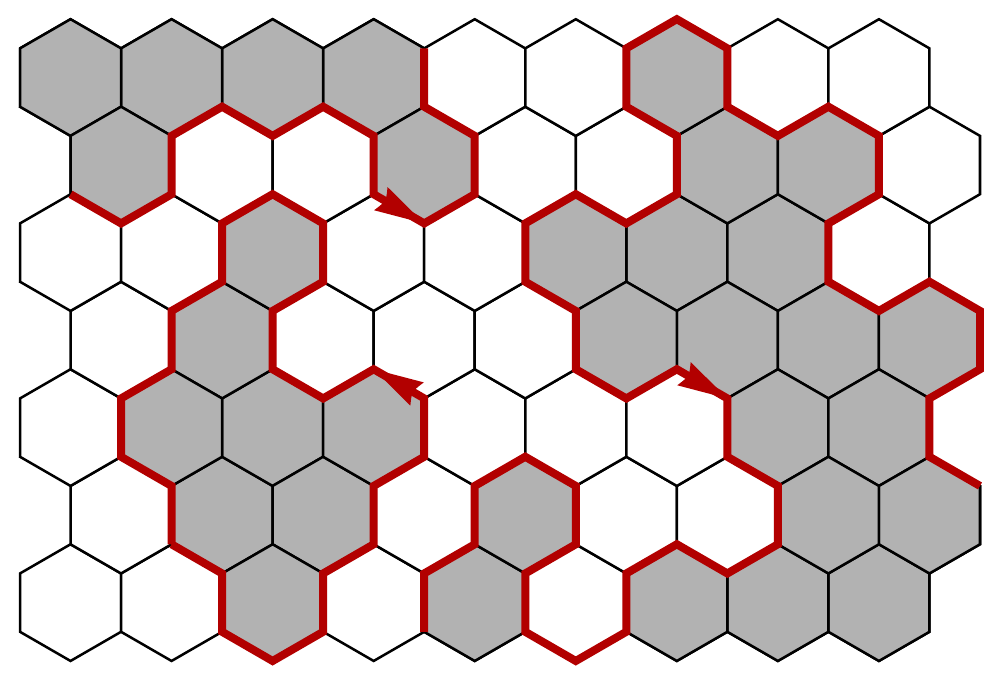}
\caption{(Color online) Percolation cluster boundaries. We will view the black clusters as surfaces in the graphical expansion of a lattice gauge theory.}
\label{percolation cluster boundaries}
\end{figure}

\noindent
We now turn to a different topic, namely 2D percolation. It is interesting to see how the relation to a lattice gauge theory is modified for two dimensional deterministic walks in a random environment in the universality class of percolation hulls (cluster boundaries).  Essentially the idea will be to take the black clusters in percolation (Fig.~\ref{percolation cluster boundaries}) to be sheets of random surface in the graphical expansion of a lattice gauge theory. This can be compared with the treatment of tricolour percolation in Sec.~\ref{tricolour percolation section}, where the random surfaces generated by a lattice gauge theory were interpreted as domain walls.

The end result in 2D is again a $CP^{k|k}$ model, but this time with a topological $\theta$ term. This is a known continuum representation of percolation \cite{read saleur, candu et al}, derived previously by mapping bond percolation to a supersymmetric spin chain. The present approach makes the appearance of the $\theta$ term transparent. There is also closely related work mapping (non-SUSY) lattice $CP^{n-1}$ models with a $\theta$ term to solid-on-solid models on the honeycomb lattice \cite{Affleck CPn-1 and SOS}. That work considers a graphical expansion in terms of loops very similar to the one described below. However, since it is not concerned with percolation, it uses a more conventional Boltzmann weight for the gauge field which does not yield the simple expansion in terms of percolation configurations that we desire.

\sss{Lattice gauge theory for 2D loops}

\noindent
The heuristic argument of Sec.~\ref{overview} can be adapted to the 2D case, but it is just as easy to give a concrete lattice construction. For definiteness, let's consider site percolation on the triangular lattice. The unit cells of the triangular lattice are faces of the honeycomb lattice, which are coloured black or white with probabilities $p$ and $1-p$;  the critical point is $p=1/2$. We orient the cluster boundaries (hulls) so as to have the black hexagons to the left -- see Fig.~\ref{percolation cluster boundaries}.

As for tricolour percolation (Sec.~\ref{tricolour percolation section}), we introduce supervectors $\psib_i=(z_i^1,...,z_i^{n+k},\chi_i^1,...,\chi_i^k)$ on the sites $i$ of the honeycomb lattice (now with arbitrary numbers of bosonic and fermionic components and $\psib^\dag \psib=n$) and gauge degrees of freedom $U_{ij}$ on the links. We also introduce a fugacity $x$ per unit length of loop, with $x=1$ for uncorrelated percolation. Letting $F$ denote a hexagon, the partition function we need is
\be
\notag
Z = \Tr \prod_{F}\bigg[ (1-p) + p \prod_{\<ij\>\in F} U_{ij} \bigg]
\prod_{\<ij\>} \left[ 1 + x U_{ij}\psib_i^\dag \psib_j + \text{cc} \right].
\ee
In the products over the links in a hexagon, the links are oriented anticlockwise. To be definite, we consider a finite lattice (made up of complete hexagons) with the topology of the disc. 

Now consider two different approaches to the graphical expansion of this partition function. In the first, we begin by expanding out the first product, representing the terms via the rule that a hexagon is coloured white if we take the first term and black if we take the second. Each term then corresponds to a percolation configuration $\mathcal P$,  and comes with a product of $U$s along the links of the cluster boundaries. Fixing $\mathcal{P}$, we then expand the second product (over links). Only one term survives after tracing over $U$, namely that which cancels the $U$s on the on the cluster boundaries, and this term comes with $\psib$ `worldlines' which encircle the black clusters anticlockwise. The trace over $\psib$ gives the loops a fugacity:
\be
\notag
Z = \sum_{\mathcal{P}} p^{B} (1-p)^{W}n^{\text{no. loops}}x^\text{total loop length}.
\ee
Here $B$ and $W$ are the numbers of black and white hexagons in the configuration $\mathcal{P}$.  Note that our boundary conditions mean that the region outside the lattice is considered to be white for the purposes of drawing the cluster boundaries. Percolation is the case $n=1$, and by varying $n$ at $p=1/2$ we can consider the well-known $O(n)$ loop models for various $n$ \cite{domany, cardy book, Affleck CPn-1 and SOS}.

For the second approach, we expand out only the second product in $Z$, and represent the terms by drawing in oriented links $\<ji\>$ of the lattice for each $U_{ij}\psib^\dag_i \psib_j$ we take. Tracing over $\psib$, only configurations of  mutually avoiding closed oriented loops survive (see also \cite{Affleck CPn-1 and SOS}). Then for each loop configuration $\mathcal{C}$ we trace over $U$ to get a Wilson loop:
\ba
Z&\propto \sum_{\mathcal{C}} n^{|\mathcal{C}|} W(\mathcal{C}), &
W(\mathcal{C})&=\< \prod_{\text{links } \in \mathcal{C} } U \>_U.
\end{align}
The Wilson loop $W(\mathcal{C})$, which is evaluated using the Boltzmann weight $\prod_{F}( (1-p) + p \prod_{\<ij\>\in F} U_{ij} )$ for the gauge field, has simple properties. It is zero if $\mathcal{C}$ is not equivalent to a percolation configuration: loops must alternate in their orientations (along any line) so as to allow the hexagons to be consistently coloured. Then, $W(\mathcal{C})\propto(1-p)^{W}p^{B}$. There is exponential suppression of black areas (regions encircled anticlockwise) when $p<1/2$, and of white areas (regions encircled clockwise) when $p>1/2$. When $p=1/2$, this exponential suppression disappears. 

These properties suggest that the continuum action for the gauge field $A$ will have a topological term $i({\theta}/{2\pi}) \int \dd^2 x\, E$, where $E= \epsilon_{\mu\nu}\partial_\mu A_\nu$, and that $\theta=\pi$ corresponds to $p=1/2$. Particles coupled to a gauge field at this value of $\theta$ are known as `half-asymptotic': in a $(1+1)$D quantum language, they can be separated at no energy cost, so long as they alternate (particle/antiparticle) in space \cite{Affleck CPn-1 and SOS, shankar murthy}. An illuminating discussion of half-asymptotic particles is given by Shankar and Murthy in \cite{shankar murthy}, who also mention a worldline interpretation pertinent to our considerations here.

This topological action in fact arises from the naive continuum limit of the Boltzmann  weight for our lattice gauge theory. The product of the $U$s around a hexagon   gives the flux through that hexagon,
\be
\prod_{\<ij\>\in F}U_{ij} \sim \exp \lf i \int_F \dd^2 x \, E \ri.
\ee
Using this, a formal expansion in $E$ turns the part of the Boltzmann weight dependent only on $E$ into
\be
\exp \int \dd^2 x \bigg(  i (\theta/2\pi) E  - {p(1-p)} E^2/2 + \ldots \bigg),
\ee
with $\theta= 2 p \pi$.  Since $E$ is not small, we should be suspicious of these values of the couplings. However, the fact that $\theta=\pi$ when $p=1/2$ is robust, because it reflects an additional symmetry which is present at $p=1/2$, where the loop ensemble is invariant under switching black and white hexagons. Neglecting boundary effects, this exchange equates to reversing the orientation of each loop, or equivalently to complex conjugation of $\psib$ and $\psib^\dag$ (note the necessary convention is $\chi^{**} = - \chi$ for the fermions \cite{Efetov}). This reversal of the charge of $\psib$ is equivalent to a change in the sign of $\theta$ in the continuum description  below. Thus the value of $\theta$ which corresponds to $p=1/2$ must be physically equivalent to $-\theta$, up to boundary effects; this is the case for $\theta=\pi$.

Taking account of the symmetries, and dropping higher terms, the continuum action is
\ba
\label{lagrangian for percolation}
\mathcal{L}=i(\theta/2\pi) E + |(\nabla - i A) \psib|^2 + \mu |\psib|^2 + \lambda |\psib|^4,
\end{align}
with $(\theta-\pi)\sim (p-1/2)$. 

Let us briefly discuss the phase structure of this theory when $n=1$ and $p=1/2$, under the assumption that $\lambda>0$. In order to vary $\mu$, we introduce an Ising coupling of strength $J$ between the colours of the hexagons, setting $x = e^{-2J}$ in the partition function for the lattice gauge theory. For strong Ising coupling, $J < J_c$, domain walls are suppressed and and $\psib$ is a massive field. Decreasing $J$, we pass through the Ising transition, at which $\psib$ condenses; since this is a thermodynamic transition as well as a geometrical one, it is nontrivial even when we remove the fermions by setting $k=0$, leaving a single complex scalar $z$ \cite{Affleck CPn-1 and SOS}. (It is interesting that the Ising transition can be described either in terms of a single boson $z$ coupled to the $\theta=\pi$ gauge field or -- as we can deduce from Eq.~(\ref{lagrangian for percolation}) by using bosonic duality and then the bosonization rules -- as a single Dirac fermion coupled to a $\theta=\pi$ gauge field; the Ising transition in the latter theory has been discussed in \cite{shankar murthy}.) Finally, for weak Ising coupling, $J < J_c$, we flow to the percolation critical point. In this regime domain walls proliferate and $\psib$ is condensed. Since $\psib$ is condensed, it is natural here to use a sigma model with a fixed length for $\psib^\dag \psib$, as in \cite{read saleur, candu et al}.

\section{Summary}
\label{Summary}

\noindent
Line-like topological defects are ubiquitous in three-dimensional systems, and continue to be of theoretical interest; for example, an exciting recent development is the realization that in certain quantum systems vortices can sustain topologically protected zero-energy states \cite{teo kane}. Here we have addressed the universal fractal geometry of vortices in disordered systems, a topic that has been studied numerically in diverse contexts but has lacked a field-theoretic description. By mapping a lattice model for vortices in a random complex field to a lattice gauge theory, and also via a continuum treatment based on XY duality, we related geometrical correlation functions for vortices to correlation functions in the $CP^{k|k}$ model, and explored simple consequences of this correspondence. For unoriented line defects, such as vortices in nematic fields, we argued that the appropriate field theory is the $RP^{2l|2l}$ model. These models play an important role in the classification of universality classes of geometrical critical phenomena in three dimensions.

We also argued that similar mappings can be fruitfully applied to 2D random curve ensembles, such as contour lines in a random height field, for which a lattice regularization is provided by site percolation and which are described by $CP^{k|k}$ with a $\theta$ term. Future work will apply similar ideas to other random curve problems in 2D.

Although we have focussed on vortices, we know from the results of \cite{short loop paper} that $CP^{k|k}$ and $RP^{k|k}$ models also apply to a class of three-dimensional loop models which undergo thermodynamically trivial geometrical phase transitions, and we expect these field theories to be generic descriptions for line defects that are topologically one-dimensional (and cannot branch, or terminate at a dangling end) or for deterministic walks in a random environment such as trajectories in a Lorentz lattice gas. Note that while at first sight the loop models seem remote from vortices, configurations in one of them (that on Cardy's `L' lattice \cite{cardy class C review}) can be specified in terms of percolation configurations on two interpenetrating lattices, and thus in terms of two height fields (albeit on different lattices) taking values $\pm 1$. Heuristically, then, there is a relation to zero lines of a complex (i.e. two-component real) function here too.

It is interesting that the distinction between compact and noncompact $CP^{n-1}$ models, much discussed recently in the context of deconfined criticality \cite{NCCPn-1}, appears even in the replica limit $n\rightarrow1$. It remains to be seen whether this actually gives a new universality class for vortices, or whether the geometrical transition described by $NCCP^{k|k}$ is inevitably separate from the thermodynamic (inverted XY) transition and reduces to the universality class of the compact $CP^{k|k}$ model (Sec.~\ref{XY critical point}). On the other hand, we do believe that unoriented vortices (or loops in an appropriate loop model) will show a localized -- extended transition in a distinct universality class from that of oriented vortices -- it remains to perform detailed simulations for this universality class, and to confirm that distinct exponents are indeed obtained.

\begin{acknowledgements}

\noindent
We are grateful for helpful discussions with J. Cardy and I. Gruzberg, and for correspondence with T. Senthil. This work was supported in part by EPSRC under Grant EP/I032487/1.

\end{acknowledgements}

\end{document}